
\input harvmac
\input epsf

%

\setbox\strutbox=\hbox{\vrule height12pt depth5pt width0pt}

\def\strut{\relax\ifmmode\copy\strutbox\else\unhcopy\strutbox\fi}

\nref\rzama{A.B. Zamolodchikov, Higher-order integrals of motion in
two-dimensional models of field theory with a broken conformal
symmetry, JETP Lett. 46 (1987) 160.}

\nref\rzamb{A.B. Zamolodchikov, Integrals of motion in scaling 3-state
Potts model field theory, Int. J. Mod. Phys. A3 (1988) 743.}

\nref\rzamc{A.B. Zamolodchikov, Integrable field theory from conformal
field theory, Advanced Studies in Pure Math. 19 (1989),641.}
\nref\rabf{G.E. Andrews, R.J. Baxter and P.J. Forrester, Eight-vertex
SOS model and generalized Rogers-Ramanujan-type identities,
J. Stat. Phys. 35 (1984) 193.}

\nref\rfb{P.J. Forrester and R.J. Baxter, Further exact solution of
the eight-vertex SOS model and generalizations of the Rogers-Ramanujan
identities, J. Stat. Phys. 38 (1985) 435.}

\nref\rfour{C.N. Yang and C.P. Yang, One-dimensional chain of
anisotropic spin-spin interactions.I Proof of Bethe's hypothesis for
ground state in a finite system, Phys. Rev. 150 (1966) 321;
One-dimensional chain of anisotropic spin-spin
interactions. II. Properties of the ground-state energy per lattice
site for an infinite system, Phys. Rev. 150 (1966) 327}

\nref\rsix{R.J. Baxter, Eight -vertex model in lattice statistics, 
Phys. Rev. Lett. 26 (1971) 832;
One-dimensional anisotropic Heisenberg model, Phys. Rev. Lett. 26 (1971)
834.}
\nref\rfive{M.Takahashi and M. Suzuki, One dimensional anisotropic
Heisenberg model at finite temperatures, 
Prog. Theo. Phys. 48 (1992) 2187.}
\nref\rnine{ A. Leclair, Restricted Sine Gordon theory and the minimal
conformal series, Phys. Lett. B 230 (1989) 103.}
\nref\rten{F.A. Smirnov, Reductions of quantum sine-Gordon model as
perturbations of minimal models of conformal field theory,
Nucl. Phys. B337 (1990) 156.}
\nref\rzamtba{Al.B. Zamolodchikov, Thermodynamic Bethe Ansatz for
RSOS scattering theories, Nucl. Phys. B358 (1991) 497;
From tricritical Ising to critical Ising by thermodynamic Bethe
Ansatz, Nucl. Phys. B358 (1991)524.}
\nref\rlz{S. Lukyanov and A.B. Zamolodchikov, Exact expectation values
of local fields in the quantum sine Gordon model,
Nucl. Phys. B 493 (1997) 571.}
\nref\rlykk{V. Fateev, S. Lukyanov, A.B. Zamolodchikov and Al. B
Zamolodchikov, Expectation values of local fields in Bullough-Dodd
model and integrable perturbed conformal field theories, hepth 9709034}
\nref\rfourteen{A.G.Izergin and V. Korepin, The inverse scattering
method approach to the quantum Shabat-Mikhailov model,
Comm. Math. Phys. 79 (1981) 303}
\nref\rfiveteen{R.K. Dodd and R.K. Bullough, Polynomial conserved
identities for the sine-Gordon equations, Proc. Roy. Soc. Lond. A352
(1977) 481.}
\nref\rsixteen{A.V. Zhiber and A.B. Shabat, Klein-Gordon equations
with a nontrivial group, Sov. Phys. Dokl. 24 (1979) 607;
A.V. Mikhailov, Pis'ma Zh. Eksp. Theor.Fiz. 30 (1979) 443.}
\nref\rwns{S.O. Warnaar, B. Nienhius and K. Seton, New construction of
solvable lattice models including an Ising model in a field,
Phys. Rev. Lett. 69 (1992) 710; A critical ising model in a magnetic field,
 Int. J. Mod. Phys. B7 (1993) 3727.}
\nref\rroc{Ph. Roche, On the construction of integrable dilute ADE models,
Phys. Lett. B285 (1992) 49}
\nref\rwpsn{S.O. Warnaar, P.A. Pearce,K.A. Seaton, and B. Nienhuis, 
Order parameters of the dilute A models,
J. Stat. Phys. 74 (1994) 469.}
\nref\rnineteen{A. Kuniba, Exact solution of solid-on-solid models for
twisted affine Lie algebras $A^{(2)}_{2n}$ and $A^{(2)}_{2n-1},$ Nucl. Phys. B355 (1991) 801}
\nref\rsmir{F.A. Smirnov, Exact S-matrices for $\phi_{1,2}$ perturbed
minimal models of conformal field theory, Int. J. Mod. Phys. A6 (1991) 1407.}
\nref\rckm{F. Colomo, A. Koubek and G, Mussardo, The subleading
magnetic deformation of the tricritical Ising model in two dimensions
as RSOS restrictions of the Izergin-Korepin model, Phys. Lett. B274
(1992) 367; On the S matrix of the subleading magnetic deformation of
the tricritical Ising model in two dimensions,
Int. J. Mod. Phys. A7 (1992) 5281.}
\nref\rkmm{A. Koubek, M.J. Martins and G. Mussardo, Scattering
matrices for $\phi_{1,2}$ perturbed conformal models in absence of kink
states, Nucl. Phys. B 368 (1992) 591.}
\nref\rkou{A. Koubek, S matrices of $\phi_{1,2}$ perturbed minimal
models; IRF formulation and bootstrap program, 
Int. J. Mod. Phys. A9 (1994) 1909}
\nref\relbaz{R.M. Ellem and V.V. Bazhanov, Thermodynamic Bethe Ansatz
for the subleading magnetic perturbation of the tricritical Ising model, hepth 9703026}
\nref\rzame{Al. B. Zamolodchikov, Thermodynamic Bethe Ansatz in
relativistic models: scaling 3-state Potts and Lee-Yang models,
Nucl. Phys. B342 (1990) 695}
\nref\rtwentyfour{M.J. Martins, Exact resonance ADE S-matrices and
their renormalization group trajectories, Nucl. Phys. B394 (1993) 339.}
\nref\rnewrav{D. Fioravanti, F. Ravanini and M. Stanishkov,
Generalized KdV and quantum inverse scattering description of
conformal minimal models,
Phys. Lett. B367 (1996) 113.}
\nref\rrav{F. Ravanini, M. Stanishkov and R. Tateo, Integrable
perturbation of CFT with complex parameter: the $M(3,5)$ model and its
generalizations, Int. J. Mod. Phys. A11 (1996) 677.}
\nref\rtwentysix{G. Takacs, A New RSOS restriction of the
Zhiber-Mikhailov-Shabat model and $\phi_{1,5}$ perturbations of
nonunitary minimal models, Nucl. Phys. B489 (1997) 532;
H. Kausch, G. Takacs and G. Watts, On the relation of $\phi_{1,2}$ and
$\phi_{1,5}$ perturbed models and unitarity, Nucl. Phys. B489 (1997) 557.}

\nref\rtwentyseven{V.V Bazhanov, B. Nienhuis and S.O. Warnaar, Lattice
Ising model in a field: $E_8$ scattering theory,
Phys. Lett. B 322 (1994) 198.}

\nref\rcpott{G. Albertini, B.M.McCoy and J.H.H. Perk, Eigenvalue
spectrum of the superintegrable chiral Potts model, Adv. Studies in
Pure Math. vol. 19 eds. M. Jimbo, T. Miwa and A. Tsuchiya,
(Kinokuniya-Academic Press,Tokyo 1989).; Phys. Lett. A 135 (1989) 741.} 

\nref\rkkmma{R. Kedem, T.R. Klassen, B.M. McCoy and E. Melzer,
Fermionic quasi-particle representations for characters of $(G^{(1)})_1\times(G^{(1)})_1/(G^{(1)})_2,$
Phys. Letts. B 304 (1993) 263.}

\nref\rkkmmb{R. Kedem, T.R. Klassen, B.M. McCoy and E. Melzer,
Fermionic sum representations for conformal field theory characters,
Phys. Letts. B 307 (1993) 68.}

\nref\rroca{A. Rocha-Caridi, in {\it Vertex operators in mathematics
and physics}, J. Lepowsky et al eds. (Springer, Berlin 1985) p.451.}

\nref\rber{A. Berkovich, Fermionic counting of RSOS states and Virasoro
character formulas for the unitary minimal series $M(\nu, \nu+1),$
Nucl. Phys. B 431 (1994) 315.}
\nref\rwar{S.O. Warnaar, Fermionic solution of the
Andrews-Baxter-Forrester model I. Unification of the TBA and CTM methods, 
J. Stat. Phys. 82 (1996) 657.}

\nref\rbm{A. Berkovich and B.M. McCoy, Continued fractions and
fermionic representations for characters of $M(p,p')$ minimal models, 
Lett. Math. Phys. 37 (1996) 49.}

\nref\rbms{A. Berkovich, B.M. McCoy and A. Schilling,
Rogers-Schur-Ramanujan type identities for the $M(p,p')$ minimal
models of conformal field theory,
Comm. Math. Phys. (in press), q-alg-9607020.}

\nref\raba{G. Andrews and A. Berkovich, Comm. Math. Phys. (in
press),q-alg 9702008.}
\nref\rabb{G. Andrews and A. Berkovich. (in preparation)}
\nref\rnewole{S.O. Warnaar, Fermionic solution of the
Andrews-Baxter-Forrester model II: Proof of Melzer's polynomial
identities, J. Stat. Phys. 84 (1996) 49}
\nref\rnewolet{S.O.Warnaar, A note on the trinomial
analogue of Bailey's lemma, q-alg 9702021.}

\nref\rbail{W.N. Bailey, Some identities in combinatory analysis 
Proc. London Math. Soc. (2) 49 (1947) 421; Identities of the
Rogers-Ramanujan type,
Proc. London Math. Soc. (2) 50 (1949) 1.}

\nref\rab{G.E. Andrews and R.J. Baxter, Lattice gas generalizations
of the hard-hexagon model III. $q$-trinomial coefficients, 
J. Stat. Phys. 47 (1987) 297.}
\nref\rscha{A. Schilling, Polynomial fermionic forms for the branching
functions of the rational coset conformal field theories 
$({\widehat su}(2))_M\times({\widehat su}(2))_N/({\widehat su}(2))_{M+N},$
Nucl. Phys. B 459 (1996) 393.}
\nref\rschb{A. Schilling, Multinomial and polynomial bosonic forms for
the branching functions of the 
$({\widehat su}(2))_M\times({\widehat su}(2))_N/({\widehat su}(2))_{M+N}$
conformal coset models, Nucl. Phys. B 467 (1996) 247.}
\nref\rmel{T.R. Klassen and E. Melzer, The thermodynamics of purely
elastic scattering theories and conformal perturbation theory,
Nucl. Phys. B350 (1991) 635.}
\nref\rmus{G. Mussardo, Off-critical statistical models: factorized
scatering theories and bootstrap program, Phys. Repts. 218 (1992) 251.}
\nref\rfat{V.A. Fateev, The exact relations between the coupling
constants and the masses of particles for the integrable perturbed
conformal field theories, Phys. Letts. B 324 (1994) 45.}

\nref\rsw{A. Schilling and S.O.Warnaar, Supernomial coefficients,
polynomial identities and $q$-series, q-alg 9701007.}

\nref\rbmo{A. Berkovich, B.M. McCoy and W.P. Orrick, 
Polynomial identities, indices, and duality for the $N=1$
supersymmetric models $SM(2,4\nu),$ J. Stat. Phys. 83 (1996) 795.}
\nref\rnewand{G.E. Andrews, $q$-Trinomial coefficients and
Rogers-Ramanujan type identities, in {\it Analytic Number Theory},
B. Berndt et al eds. (Boston, Birkha{\"u}ser (1990)) 1.}

\nref\rmac{P.A. MacMahon, {\it Combinatory Analysis,} vol. 2
(Cambridge University Press, Cambridge 1916)}

\nref\rschur{I.J. Schur, Ein Beitrag zur additiven Zahlentheorie und
zur Theorie der Kettenbr{\" u}che, S-B. Preuss,
Akad. Wiss. Phys.-Math. Kl. (1917) 302.}
\nref\randrews{G.E. Andrews, A polynomial identity which implies the
Rogers-Ramanujan identities, Scripta Mathematica. 28 (1970) 297.}
\nref\rburge{W.H. Burge, Restricted partition identities, 
J. Comb. Theor., A63 (1993) 210.}

\nref\romar{O. Foda, K.S. Lee and T.A. Welsch,
A Burge tree of Virasoro-type polynomial identities, q-alg 9710025.}

\nref\rjkm{J.D. Johnson, S. Krinsky and B.M. McCoy, 
Vertical-arrow correlation length in the eight-vertex model and the
low-lying excitations of the X-Y-Z Hamiltonian, Phys. Rev. A8
(1973) 2526.}
\nref\rluth{A. Luther, Eigenvalue spectrum of interacting massive
fermions in one dimension, Phys. Rev. B14 (1976) 2153.}
\nref\rlush{M. L{\"u}scher, Dynamical charges in the quantized
renormalized massive Thirring model, 
Nucl. Phys. B117 (1976) 475.}
\nref\rjnw{G.I. Japaridze, A.A. Nersesyan, and P.B. Wiegmann, Exact
results in the two-dimensional $U(1)$ symmetric Thirring model,
Nucl. Phys. B230[FS10] (1984) 511.}
\nref\rkor{V.E. Korepin, The mass spectrum and the S matrix of the
massive Thirring model in the repulsive case, 
Comm. Math. Phys. 76 (1980) 165.}

\nref\rik{A.G. Izergin and V. Korepin, The lattice quantum sine-Gordon
equation, Lett. Math. Phys. 5 (1981) 199.}
\nref\rbk{N.M. Bogolyubov and V.E. Korepin, Structure of the vacuum in
the quantum sine-Gordon model, Phys. Lett. B 159 (1985) 345.}

\Title{\vbox{\baselineskip12pt
  \hbox{ITPSB 97-70}}}
  {\vbox{\centerline{The perturbations ${\phi_{2,1}}$ and ${\phi_{1,5}}$}
\centerline{  of the minimal models $M(p,p')$}
\centerline{ and the trinomial analogue of Bailey's lemma} }}


  \centerline{ Alexander Berkovich\foot{alexb@insti.physics.sunysb.edu} 
               and Barry M. McCoy\foot{mccoy@max.physics.sunysb.edu}}
\bigskip\centerline{\it Institute for Theoretical Physics}
  \centerline{\it State University of New York}
  \centerline{\it Stony Brook,  NY 11794-3840}
  \bigskip
\centerline{Paul A. Pearce\foot{pap@maths.mu.oz.au}}
\bigskip\centerline{\it Department of Mathematics and Statistics}
\centerline{\it University of Melbourne}
\centerline{\it Parkville, Victoria 3052, Australia}
  \Date{\hfill 12/97}

  \eject

\centerline{\bf Abstract}
We derive the fermionic polynomial generalizations of the
characters of the integrable perturbations
$\phi_{2,1}$ and $\phi_{1,5}$ of the 
general minimal $M(p,p')$ conformal
field theory by use of the recently discovered trinomial analogue of
Bailey's lemma.
For $\phi_{2,1}$ perturbations  results are given for 
all models with $2p>p'$ and for $\phi_{1,5}$ perturbations results for
all models with ${p'\over 3}<p< {p'\over 2}$ are obtained. 
For the $\phi_{2,1}$ perturbation of the unitary case $M(p,p+1)$ 
we use the incidence matrix obtained
from these character polynomials to conjecture  a set of TBA
equations.  
We also find that for $\phi_{1,5}$ with $2<p'/p < 5/2$ and 
for $\phi_{2,1}$ satisfying $3p<2p'$ there
are usually several different fermionic polynomials which lead to the
identical bosonic polynomial. We interpret this to mean that in these
cases the specification of the perturbing field is not sufficient to
define the theory and that an independent statement of the choice of
the proper vacuum must be made.

\newsec{Introduction}

The theory of perturbations of conformal field theory was
initiated by Zamolodchikov \rzama-\rzamc~ who introduced a 
sufficient condition, known as the ``counting argument'',
for an operator to provide an integrable perturbation of the minimal
conformal field theories $M(p,p').$
It is well known that this argument indicates that
the operators $\phi_{1,3},~\phi_{1,2},~\phi_{2,1}$ and $\phi_{1,5}$ are
integrable whenever they are relevant. 
Of these perturbations the $\phi_{1,3}$ is the best understood and is
known to be related to the affine Lie algebra $A^{(1)}_1$. On the one
hand it is realized as the scaling limit of RSOS
lattice models \rabf-\rfb~which are restrictions of the XXZ \rfour~
and XYZ \rsix-\rfive~ spin chains.
In the field theory context it is related
to restrictions of the sine-Gordon model \rnine-\rten. The scattering
matrices \rnine-\rten, Thermodynamic Bethe's Ansatz (TBA) equations
\rzamtba~
and the vacuum expectation values of some local operators are known \rlz-\rlykk. In all studies there
is agreement with the perturbative treatment and there is a sense in
which the perturbation may be said to define the model.

The other three perturbations, $\phi_{2,1},~\phi_{1,2},$ and
$\phi_{1,5},$ are not nearly so well understood. They
are known to be related to the affine Lie algebra $A^{(2)}_2$ and to
the Izergin-Korepin lattice model \rfourteen. In the field theory
context they are related to the Bullough-Dodd\rfiveteen 
/Zhiber-Mikhailov-Shabat\rsixteen
model. In contrast to the unique way to restrict the 
XXZ model there are two different restrictions for the Izergin-Korepin model. One is the
restriction known as the dilute $A_L$ model of \rwns-\rwpsn~
 and the other is
the restriction due to Kuniba \rnineteen. In principle these restrictions
can be studied for all models $M(p,p')$ but in practice the unitary
case $M(p,p+1)$ has been investigated the most. In field theory the S
matrices for these perturbations have been determined \rsmir-\rkou~
and TBA equations have been obtained for the following cases: The
$\phi_{2,1}$ perturbation of $M(4,5)$ \relbaz~ and of  $M(5,6)$
\rzame, the
$\phi_{2,1}$ perturbation of $M(p,2p-1)$ and the $\phi_{1,5}$
perturbation of $M(p,2p+1)$ of \rtwentyfour-\rrav,
 the $\phi_{1,5}$ perturbation of
$M(3,10),~M(3,14),~M(3,16)$ by \rtwentysix~ and the $\phi_{1,2}$ perturbation
of $M(3,4)$ \rtwentyseven.

One of the reasons that the results for the $\phi_{2,1},~\phi_{1,2}$
and $\phi_{1,5}$ perturbations are not nearly so extensive as for
$\phi_{1,3}$ is that there appear to be new physical effects which are
not present for the $\phi_{1,3}$ perturbation. These effects are seen
vividly in the recent study of vacuum expectation values
by Fateev, Lukyanov, Zamolodchikov and
Zamolodchikov \rlykk. In particular they find for the
$\phi_{1,5}$ perturbation of the minimal model $M(p,p')$ that in the
region $2<p'/p\leq 8/3$ the mass defined by perturbation theory is
negative. This indicate that a non perturbative definition of the
vacuum must be given and is often a signal that level crossing of the
vacuum has occurred \rcpott . Similarly for the $\phi_{2,1}$ perturbation
they find a problem with vacuum definition for $3p>2p'.$ None of these
problems has been found to occur for the $\phi_{1,3}$ perturbation.

In this paper we study the $\phi_{2,1}$ and the $\phi_{1,5}$
perturbations by the method of fermionic representations of Virasoro
characters and Rogers-Ramanujan identities. This
approach gives results for $\phi_{2,1}$ for all values of $p'/p$ and
for $\phi_{1,5}$ for $2<p'/p<3$ and thus gives results even  
in the regime where the perturbative
definition of the field theory leads to problems of vacuum choice.

This method began 
several years after the work of \rzama-\rzamc~when it was proposed 
\rkkmma-\rkkmmb~ that integrable perturbations are closely connected to
the various different fermionic representations of the characters of
the model. For example, for the minimal models the (bosonic)
representation of the characters $\chi_{r,s}^{(p,p')}(q)$ (normalized to
$1$ at $q=0$) is~\rroca
\eqn\roch{\chi_{r,s}^{(p,p')}(q)={1\over
(q)_{\infty}}\sum_{j=-\infty}^{\infty}\left(q^{j(p p' j +p' r
-ps)}-q^{(pj+r)(p' j+s)}\right)}
where
\eqn\qdef{(a)_n=\cases{\prod_{j=0}^{n-1}(1-aq^j)&for $n\geq 1$\cr
1&for $n=0.$\cr}}
which have central charge 
\eqn\cen{c=1-{6(p-p')^2\over p p'}}
and conformal dimensions
\eqn\dim{\Delta_{r,s}={(ps-p'r)^2-(p-p')^2\over 4p p'}}
where $p$ and $p'$ are relatively prime integers with $p<p'.$
It was conjectured in \rkkmmb~and proven in~\rber-\rwar~
that for $M(p,p+1)$ the character $\chi_{1,1}^{(p,p+1)}(q)$
has the fermionic representation
\eqn\ferot{\sum_{{\bf m}\atop m_{p-2}\equiv 0~({\rm mod}2)}
q^{{1\over 4}{\bf mC}_{p-2}{\bf m}}{1\over (q)_{m_1}}\prod_{j=2}^{p-2}
{({1\over 2}{\bf I}_{p-2}{\bf m})_j\atopwithdelims[] m_j}_q=
\chi_{1,1}^{(p,p+1)}(q)}
where the $p \times p$ dimensional matrices
\eqn\in{({\bf I}_p)_{j,k}=\delta_{j,k-1}+\delta_{j,k+1}~~{\rm and}~~
{\bf C}_p=2-{\bf I}_p,~~{\rm for}~1\leq i,k\leq p}
are the incidence and Cartan matrix (respectively) of the Lie algebra $A_p$
and the $q$-binomials are defined by
\eqn\qbin{{m+n \atopwithdelims[] m}_q=\cases{{(q)_{m+n}\over
(q)_m(q)_n}&for $m,n$ nonnegative integers\cr
0& otherwise.\cr}}
The sum in ~\ferot~is defined to be the sum over all integers values
of $m_i$ with $1\leq i\leq p-2$ such that ~\qbin~is not 
zero and the component $m_{p-2}$ has
the additional restriction that $m_{p-2}\equiv~0~({\rm mod}2).$ These
conventions will be followed in all sums in this paper.
Since the incidence matrix ${\bf I}_{p-2}$ is an important ingredient
in the Thermodynamic Bethe Ansatz (TBA) equations for the
$\phi_{1,3}$ perturbation of $M(p,p+1)$ models ~\rzamtba~it is natural
to associate fermionic representation~\ferot~with the perturbation
$\phi_{1,3}$

This identification of integrable perturbations with
fermionic representations may be extended to the polynomial
generalizations of the characters which are used to prove the identity
of the fermionic and bosonic forms of the characters.
As an example, the polynomial generalization of ~\roch~
used in ~\rber~ to prove~\ferot~for the minimal conformal dimension 
~\dim~with $|ps_m-p' r_m|=1$ and $L$ is even is the path counting formula 
of~\rabf-\rfb~
\eqn\pboseot{B(L,p,p')=\sum_{j=-\infty}^{\infty}\left(q^{j(p
p'+1)}{L \atopwithdelims[] {L\over 2}-p'j}_q-q^{(pj+r_m)(p'
j+s_m)}{L\atopwithdelims[] {L\over 2}-jp'-s_m}_q\right)}
which if we use
\eqn\limqbin{\lim_{L\rightarrow \infty}{L\atopwithdelims[] {L\over
2}-a}_q={1\over (q)_{\infty}}} reduces to ~\roch~(with $r=r_m,~s=s_m$)
as $L\rightarrow \infty.$
Similarly the polynomial generalization of the fermionic form in ~\ferot~is
\eqn\pfermot{F(L,p,p+1)=
\sum_{{\bf m}\atop{ m_{p-2}\equiv 0({\rm mod}2)}}
q^{{1\over 4}{\bf mC}_{p-2}{\bf m}}\prod_{j=1}^{p-2}{({1\over 2}{\bf
I}_{p-2}{\bf m}+{L\over 2}{\bf e_1})_j\atopwithdelims[] m_j}_q}
where we use the definition of the $p-2$ dimensional
unit vectors $({\bf e}_j)_k=\delta_{j,k}.$
In the rest of this paper we will use ${\bf e}_j$ for the $j^{th}$
unit vector and assume that its dimension is clear from the context.
The polynomial ~\pfermot~reduces to the left hand side of ~\ferot~in the
limit $L\rightarrow \infty$ by use of
\eqn\xnew{\lim_{L\rightarrow \infty}{L\atopwithdelims[] m}_q={1\over (q)_m}}
The character identity ~\ferot~ itself thus generalizes to the
polynomial identity
\eqn\polyot{F(L,p,p+1)=B(L,p,p+1).}
Polynomial analogues of Rogers-Ramanujan type identities 
have been obtained for all
models $M(p,p')$ \rbm, \rbms~ and for all models where independent
evidence exists \rabf-\rfb,\rzamtba~ 
these identities yield the particle content and are
related to the TBA equations for the perturbation $\phi_{1,3}.$

The purpose of this paper is to extend these Rogers-Ramanujan studies
from the $\phi_{1,3}$ perturbations~\rber-\rbms~ to
the perturbations $\phi_{2,1}$
and $\phi_{1,5}$ of the general models $M(p,p').$ For the $\phi_{2,1}$
perturbation of $M(p,p+1)$ 
we then use these identities to conjecture a TBA which is identical
with that of ~\relbaz~for $p=4.$

We also make  the unexpected discovery that there are certain
minimal models where one bosonic polynomial representation of the
character may have several different fermionic representations. This
is a phenomena not seen in any previous study. It seems significant
that for the perturbations $\phi_{2,1}$ and $\phi_{1,5}$ 
the condition for these
multiple representations of the polynomials to occur  
restricts $p'/p$ to lie in precisely the regions where the study of
vacuum expectation values~\rlykk~has not been carried out because of
problems in identifying the correct vacuum.
Thus our results
provide some insight into the question of vacuum choice in quantum
field theory.

Our method relies on the recently discovered \raba, \rabb~ trinomial 
analogue of Bailey's lemma plus additional results on trinomial
representations of bosonic character polynomials. 
These results are summarized in sec. 2. In sec. 3 we study the case
of $\phi_{2,1}$ perturbations of $M(p,p+1)$ and make contact with the
$\phi_{2,1}$ perturbations studied in the dilute $A_n$ models in
~\rwpsn. In sec. 4  we relate our result to
the recent TBA study~\relbaz~ of $M(4,5)$
and use this to discuss  possible TBA equations 
for the $\phi_{2,1}$ perturbation of some other unitary 
models $M(p,p+1).$ 
 In sec. 5 we consider $\phi_{1,5}$
perturbations of $M(p,3p-1)$. In sec. 6 we greatly extend our results to cover
the $\phi_{2,1}$ perturbations of almost all models 
$M(p,p')$ with $2p>p'$ and the $\phi_{1,5}$ perturbations of  most
models $M(p,p')$ with ${p'\over 3}<p<{p'\over 2}.$
There are two special cases not included in sec. 6 which must be
treated separately. In sec. 7 we discuss $M(p,2p\pm 2)$ for $p\geq 5$
odd and  from
generalizations arising from this case  we find in sec. 8
multiple fermionic representations of bosonic character polynomials
for the $\phi_{2,1}$ perturbation whenever $3p<2p'$ and for
$\phi_{1,5}$ with $2<p'/p \leq 5/2.$
In sec. 9 we cover the other special case
$M(p,2p\pm1)$ for $p\geq 3$ and make contact with the work of 
\rtwentyfour-\rrav.
We conclude in sec. 10 with a brief discussion of
our results.

\newsec{The Trinomial Bailey Lemma}

The $q$-trinomials and the trinomial Bailey's lemma are generalizations of
the well known q-binomials and the (binomial) Bailey lemma~\rbail.
The q-binomials were defined by ~\qbin~ and  obey
\eqn\qbininv{{m+n\atopwithdelims[]
m}_{q^{-1}}=q^{-mn}{m+n\atopwithdelims[] m}_q.}
For the corresponding definition of q-trinomials we follow ~\rab~and
define 
\eqn\qrtri{{L;B;q\atopwithdelims()A}_2
=\sum_{j=0}^{\infty}{q^{j(j+B)}(q)_L\over (q)_j(q)_{j+A}(q)_{L-2j-A}}
=\sum_{j=0}^{\infty}q^{j(j+B)}{L\atopwithdelims[]
2j+A}_q{2j+A\atopwithdelims[] j}_q}
and 
\eqn\qtri{T_0(L,A;q)=q^{{L^2-A^2\over
2}}{L;A;q^{-1}\atopwithdelims() A}_2.}
In the sequel we will only need the case $A=B$  and will suppress
the argument $B$ in \qrtri. We refer to these as ``round''
trinomials. We will also suppress all arguments $q$ when not needed.

The following properties of (round) q-trinomials
\eqn\syma{{L\atopwithdelims() A}_2={L\atopwithdelims() -A}_2}
\eqn\symb{T_0(L,A)=T_0(L,-A)}
\eqn\lima{\lim_{L\rightarrow\infty} {L\atopwithdelims() A}_2
={1\over (q)_{\infty}}}
\eqn\limb{\lim_{L\rightarrow\infty} T_0(L,A)=
\cases{{(-q^{1/2})_{\infty}+(q^{1/2})_{\infty}\over 2 (q)_{\infty}}&if $L-A$
is~even\cr
{(-q^{1/2})_{\infty}-(q^{1/2})_{\infty}\over 2(q)_{\infty}}
&if $L-A$ is~odd\cr}}
will be important in the sequel.

We say that the two sequences $\alpha_n$ and $\beta_n$ form a
{\bf trinomial Bailey pair} if
\eqn\tbpair{\beta_L=\sum_{j=0}^n\alpha_j{T_0(L,j)\over (q)_L.}}
Then from~\raba,\rabb~we have the following which we call the

{\bf Trinomial  Analogue of Bailey's Lemma (TABL)}

\eqn\tabl{\sum_{L=0}^{\infty}(\rho)_L(-1)^Lq^{L/2}\rho^{-L}\beta_L=
{(q/\rho)^2_{\infty}\over (q)_{\infty}(q/\rho^2)_{\infty}}
\sum_{n=0}^{\infty}{(-1)^nq^{n/2}\rho^{-n}(\rho)_n \alpha_n\over (q/\rho)_n}.}
The case $\rho=-1$ was dealt with in ~\raba~and is shown to lead to
identities for the $N=2$ supersymmetric models. In this paper we 
consider the case
 $\rho \rightarrow \infty$ where ~\tabl~reduces to
\eqn\ctabl{\sum_{L=0}^{\infty}q^{{L^2\over 2}}\beta_L={1\over (q)_{\infty}}
\sum_{n=0}^{\infty}q^{{n^2\over 2}}\alpha_n.}

\newsec{The $\phi_{2,1}$ perturbation of $M(p,p+1)$ for $p\geq 4$}

In order to apply TABL~\tabl~ we need to start with a 
trinomial Bailey pair~\tbpair. As is the case with binomial Bailey
pairs these pairs are often related to finitizations of some 
character identities. We thus begin our study by considering the polynomial
identities found by
Schilling~\rscha,\rschb~for the coset model $(A_1^{(1)})_2\times
(A_1^{(1)})_{p-3}/(A_1^{(1)})_{p-1}$
(the $N=1$ supersymmetric model $SM(p-1,p+1)$ with $p\geq 4$) 
\eqn\schw{\eqalign{&\sum_{{\bf m}\atop m_{p-2}\equiv({\rm mod}2)}q^{{1\over 4}{\bf
m}C_{p-2}{\bf m}}\prod_{i=1}^{p-2}{({1\over 2}{\bf I}_{p-2}{\bf
m}+{{\tilde L}\over 2}{\bf e}_2)_i\atopwithdelims[] { m}_i}_q\cr
&=\sum_{j=-\infty}^{\infty}\left(q^{{j\over 2}(j(p+1)(p-1)+2)}
T_0({\tilde L},(p+1)j)-q^{{1\over 2}((p+1)j+1)
((p-1)j+1)}T_0({\tilde L},(p+1)j+1)\right)\cr}}
with ${\bf I}_{p-2}$ defined in ~\in.
Together with the definition~\tbpair~ this gives the trinomial Bailey pair
\eqn\paira{\eqalign{\beta_{\tilde L}&
={1\over (q)_{{\tilde L}}}\sum_{{\bf m}\atop m_{p-2}\equiv
0({\rm mod}2)}q^{{1\over 4}{\bf
m}C_{p-2}{\bf m}}\prod_{i=1}^{p-2}{({1\over 2}{\bf I}_{p-2}{\bf
m}+{{\tilde L}\over 2}{\bf e}_2)_i\atopwithdelims[] { m}_i}_q\cr
\alpha_n&=\cases{q^{j(j(p+1)(p-1)+2)/2}&for $j\geq 0,~n=(p+1)j$\cr
q^{j(j(p+1)(p-1)-2)/2}&for $j\geq 1,~n=(p+1)j$\cr
-q^{((p+1)j+1)((p-1)j+1)/2}&for $j\geq 0,~n=(p+1)j+1$\cr
-q^{((p+1)j-1)((p-1)j-1)/2}&for $j\geq 1,~n=(p+1)j-1$\cr}}}

To this pair we now apply the special case~\ctabl~of the TABL and
obtain the identity 
\eqn\ident{\eqalign{&\sum_{{\tilde L}=0}{q^{{{\tilde L}^2\over
2}}\over(q)_{\tilde L}}\sum_{{\bf m}\atop m_{p-2}\equiv 0
({\rm mod}2)}q^{{1\over 4}{\bf
m}C_{p-2}{\bf m}}\prod_{i=1}^{p-2}{({1\over 2}{\bf I}_{p-2}{\bf
m}+{{\tilde L}\over 2}{\bf e}_2)_i\atopwithdelims[] { m}_i}_q\cr
&={1\over
(q)_{\infty}}\sum_{j=-\infty}^{\infty}\left(q^{j(jp(p+1)+1)}-\
q^{(jp+1)(j(p+1)+1)}\right)=\chi_{1,1}^{(p,p+1)}(q)}}
where in the last line  we have made the identification 
with the bosonic character formula ~\roch. However, even though the
bosonic side of this identity is the same as the bosonic side of
~\ferot ~the fermionic side is  not. In particular the sum in ~\ferot~
involves $p-2$ variables while the $q$-series in the left hand side of
\ident~ involves $p-1$
variables. Thus the TABL has produced a new fermionic representation
of the characters $\chi_{1,1}^{(p,p+1)}(q)$ which we must now 
identify with   some integrable perturbation. 

The first step in the identification of the physical meaning of the
q-series in ~\ident~ is to write it in the ``canonical quasi-particle
form''~\rkkmma,\rkkmmb~ in terms of a single (possibly asymmetric) 
matrix $B$ and a vector
$\bf u$ (which will in general have some infinite components)
\eqn\fermiform{F({\bf u})=\sum_{{\bf m}\atop {\rm restrictions}}
q^{{1\over 2}{\bf mBm}}\prod_{j}{((1-{\bf
B}){\bf m}+{\bf u})_j\atopwithdelims[] m_j}_q.}
Such a representation is indeed possible if we set 
\eqn\udef{{\tilde L}=m_0,~u_0=\infty,~~~u_i=0~~{\rm for}~~ i\neq 0}
and define  the
$p-1\times p-1$ dimensional matrices
\eqn\bdefn{{\bf B}={1\over 2}{\bf{\tilde C}}_{p-1},~~~{\rm with}~{\bf{\tilde
C}}_{p-1}=2-{\bf{\tilde I}}_{p-1}}
where
\eqn\idefn{{\bf {\tilde I}}_{p-1}=
\pmatrix{0&0&-1&0&0&\ldots&0&0&0\cr
         0&0&1&0&0&\ldots&0&0&0\cr
	 1&1&0&1&0&\ldots&0&0&0\cr
	 0&0&1&0&1&\ldots&0&0&0\cr
         0&0&0&1&0&\ldots&0&0&0\cr
         \vdots&\vdots&\vdots&\vdots&\vdots&\vdots\vdots\vdots
&\vdots&\vdots&\vdots\cr
         0&0&0&0&0&\ldots&1&0&1\cr
         0&0&0&0&0&\ldots&0&1&0\cr}}
This ``incidence matrix'' is represented graphically in Fig. 1 where 
we use the conventions that if sites $a$ and $b$ are connected by a
line then $({\bf \tilde I}_{p-1})_{a,b}=({\bf \tilde I}_{p-1})_{b,a}=1$ 
and if $a$ and $b$ are joined
by an arrow pointing from $a$ to $b$ then $({\bf \tilde  I}_{p-1})_{a,b}
=-({\bf \tilde I}_{p-1})_{b,a}=-1.$
This convention will be used throughout the entire paper.

We now need to identify this q-series identity with a perturbation of
$M(p,p+1).$ One way to do this is to consider the
polynomial generalization of ~\ident~
\eqn\polyidto{F_{2,1}(L,p,p+1)=B_1(L,p,p+1)}
where
\eqn\polyfer{F_{2,1}(L,p,p+1)=\sum_{{\bf m}\atop m_{p-2}\equiv 0 ({\rm mod}2)}
q^{{1\over 4}{\bf m}{\bf \tilde C}_{p-1}{\bf
m}}\prod_{j=0}^{p-2}{({1\over 2}{\bf \tilde I}_{p-1}{\bf m}+L{\bf
e}_0)_j\atopwithdelims[] m_j}_q.}
and
\eqn\trib{B_1(L,p,p')=\sum_{j=-\infty}^{\infty}
\left(q^{j(pp' j+r_mp'-ps_m)}{L\atopwithdelims() 2pj}_2-
q^{(jp+r_m)(jp'+s_m)}{L\atopwithdelims() 2pj+2r_m}_2\right)}
where
\eqn\rmsm{|p'r_m-ps_m|=1}
which by use of ~\lima~and \xnew~reduces to ~\ident~as $L \rightarrow \infty.$
We have verified that~\polyidto~holds by extensive computer checks.
The polynomial~\trib~for $p'=p+1$ appears in the 
calculation~\rwpsn~ of the order parameters  of
the dilute $A_p$ models in regime $1^{\pm}$ and corresponds to the
$\phi_{2,1}$ perturbations of $M(p,p+1).$ We  will here generalize
this and conclude that the trinomial generalization ~\trib~of the 
character $\chi_{r_m,s_m}^{(p,p')}(q)$ is the polynomial bosonic
representation of the $\phi_{2,1}$ perturbation of all models $M(p,p')$
whenever $2p>p'$ so that the perturbation is relevant. We also note
that an identity somewhat similar to ~\polyidto~appeared in
~\rnewole~as eqn. (5.1) and was proven in ~\rnewolet. 
The right hand side of that equation is indeed
the same as the right hand side of ~\polyidto. However, the left hand
side of the identity of ~\rnewole~is not the same as ~\polyfer. In
particular we note: (1) for finite $L$ the fermionic form of~\rnewole~
is not in the
canonical form~\fermiform~which has a quasi-particle interpretation; and
(2) in the limit $L\rightarrow \infty$ the fermionic form of
~\rnewole~is identical with ~\ferot~and not with the left hand side of ~\ident.

\newsec{A TBA conjecture for the $\phi_{2,1}$ perturbation of
$M(p,p+1)$ for $p\geq 4$}

The incidence matrix ${\bf \tilde I}_{p-1}$~\idefn~for $p=4$ has an
important connection with the TBA computations for the
model $M(4,5)$ perturbed by $\phi_{2,1}$ recently done by Ellem and 
Bazhanov~\relbaz. In this section we explain this connection and use
it to conjecture a system of TBA equations for  more general case
of the model $M(p,p+1).$

To explain the connection we rewrite the fermionic
polynomial~\polyfer~in terms of what is called an $m,n$ system as follows
\eqn\mnpoly{F_{2,1}(L,p,p+1)=\sum_{{\bf m}\atop m_{p-2}\equiv 0 ({\rm
mod}2)}q^{{1\over 4}{\bf
m}{\tilde {\bf C}}_{p-1}{\bf m}}\prod_{j=0}^{p-2}
{m_j+n_j\atopwithdelims[] m_j}}
where
\eqn\nmsy{{\bf n}+{\bf m}={1\over 2}{\bf {\tilde I}}_{p-1}{\bf m}+L{\bf
e}_0}
and $m_i$ and $n_i$ are nonnegative integers.
To make contact with ~\relbaz~we specialize to the case $p=4$ to
find
\eqn\nmpf{\eqalign{n_0+m_0&=L-{1\over 2}m_2\cr
n_2+m_2&={1\over 2}(m_0+m_1)\cr
n_1+m_1&={m_2\over 2}.\cr}}
This system is identical with the corresponding system which may
be obtained from (3.13)-(3.16) of ~\relbaz~by integrating (3.13)
of~\relbaz~ on
$\theta$ and using the notation
\eqn\baznot{\int d\theta \sigma_k(\theta)=m_k, ~{\rm and}~\int d\theta {\tilde
\sigma}_k(\theta)=n_k}
and letting
\eqn\leqn{{m\over 2 \pi}\int d\theta \cosh \theta \rightarrow L}
where we note that the sign factor $s_0=-1$ in \relbaz~corresponds to
the asymmetry present in the matrix ${\bf {\tilde  I}}_{p-1}$ of ~\idefn.

From these considerations it is natural to try to extend the TBA
equations of~\relbaz~for $M(4,5)$ to $M(p,p+1)$ by writing
the following set which reduces
to (4.4) and (4.5) of ~\relbaz~when $p=4$
\eqn\tbaa{\epsilon_j(\theta)=\delta_{j,0}r
\cosh\theta+\sum_{k=0}^{p-2}\int_{-\infty}^{\infty}
\Phi_{j,k}(\theta-\theta')\ln (1+e^{-\epsilon_k(\theta')})d\theta'
,~~j=0,1,\cdots,p-2}
with the ground state scaling function $c(r,p)$  given by
\eqn\tbab{{\pi c(r,p)\over 6r}={1\over 2\pi}\int_{-\infty}^{\infty}\cosh
\theta \ln(1+e^{-\epsilon_0(\theta)})d\theta}
where
\eqn\tbac{\Phi_{j,k}(\theta)=-\phi_0(\theta)({\bf \tilde
I}^t_{p-1})_{j,k}-\delta_{j,0}\delta_{k,0}\phi_1(\theta),}
\eqn\tbad{\phi_0(\theta)={3(p-1)\over 2 \pi \cosh 3(p-1)\theta},
~~~\phi_1(\theta)={1\over 2\pi i} {d\over d\theta} \ln F_{CDD}(\theta;p)}
where in order to assure that
 \eqn\tbaj{c(0,p)=1-{6\over p(p+1)}}
we require that the integral
$\int_{-\infty}^{\infty}\phi_{1}(\theta)d\theta$ be zero.

To complete this TBA equation requires the specification of
$F_{CDD}(\theta;p).$
In ~\relbaz~this is computed from the S-matrix of
Smirnov~\rsmir. Here, however, we will content ourselves by merely
examining the assumption that
\eqn\tbacdd{F_{CDD}(\theta;p)=\prod_{j}F_{\alpha_j}(\theta)}
where
\eqn\tbai{F_{\alpha}(\theta)={\sinh \theta+i \sin \alpha \pi\over
\sinh \theta -i \sin\alpha \pi}.}
Then, in order that as $r\rightarrow 0$ the term in $c(r,p)$ 
of order $O(r^2)$
(which is computed using the methods of~\rzamtba,\rzame,\rmel,\rmus)
agrees with the bulk energy found by Fateev~\rfat~we find that
$\alpha_j$ must satisfy the constraint
\eqn\constr{\sum_{j}\sin \pi \alpha_j={\sin {\pi\over 3}\sin{2\pi
p\over 3(p-1)}\over \sin{\pi (p+1)\over 3(p-1)}}.}
This constraint has the following simple solution  
\eqn\consol{\alpha_j=\cases{{1\over 3}+{j2(p+1)\over
3(p-1)},~j=0,\cdots, p-2&for $p\equiv 0({\rm mod 3})$\cr
{j2(p+1)\over 3(p-1)},~~j=1,\cdots,p-2&for $p\equiv 1({\rm mod}3).$\cr}}
When $p=4$ this reduces to the expression used in ~\relbaz~of
$F_{CDD}(\theta;4)=F_{-{1\over 9}}(\theta)F_{2\over 9}(\theta).$
However, for $p\equiv 2~({\rm mod}3)$ (which includes the three state
Potts model) a simple solution to the constraint~\constr~does not seem
to exist. We also note that the requirement
$\int_{-\infty}^{\infty}\phi_{1}(\theta)d\theta=0$ 
will be satisfied if (once all the $\pi\alpha_j$ are reduced mod
$2\pi$ to lie in the interval $-\pi<\pi\alpha_j<\pi$) the number of
positive and negative $\alpha_j$ are the same.  There are, however,
many cases where this does not hold and here vanishing of 
$\int_{-\infty}^{\infty}\phi_1(\theta)d\theta$ can be ensured
by the addition of delta functions to 
$\phi_1(\theta)$ which
may be thought of as a limiting case of ${1\over 2\pi i}{d\over
d\theta}{\rm ln} F_{\alpha}(\theta)$ 
as $\alpha\rightarrow 0.$ 
The case $p\equiv 2~({\rm mod} 3)$ will be discussed elsewhere.
Further tests of the equations \tbaa-\consol~ are of course 
possible but are beyond the scope of this paper.

%

\newsec{The $\phi_{1,5}$ perturbation of $M(p,3p-1)$ for $p\geq 4$}

In sec. 3 we went from the identities~\schw~for the models
$SM(p-1,p+1)$ to the identities~\polyidto~for the models $M(p,p+1).
$ Remarkably we may apply the TABL once again
if we first send $q\rightarrow 1/q$ in the polynomial identity~
\polyidto~and use ~\qtri~with $n=0$ to convert the round trinomials into $T_0$
so that a trinomial Bailey pair may be extracted using the
definition~\tbpair. Thus  
applying~\ctabl~ once again we have the new identity
\eqn\phiofident{\sum_{\tilde L=0}^{\infty}
\sum_{{\bf m}\atop m_{p-2}\equiv 0~({\rm mod}2)}
q^{{\tilde L}^2-{\tilde L}m_0
+{1\over 4}{{\bf m}{\bf \tilde C}_{p-1}{\bf m}}}{1\over
(q)_{\tilde L}}\prod_{j=0}^{p-2}{({1\over 2}{\bf \tilde I}_{p-1}{\bf
m}+{\tilde L}{\bf
e}_{0})_j\atopwithdelims[] m_j}_q=\chi_{1,3}^{(p,3p-1)}(q)}

We again want to generalize this to a polynomial which we do
by first replacing in 
~\phiofident~${\tilde L}\rightarrow n_{-1}$ and then letting
\eqn\repla{{1\over (q)_{n_{-1}}}\rightarrow 
{L-n_{-1}+m_0\atopwithdelims[] n_{-1}}_q}
to define a polynomial
\eqn\phifop{F_{1,5}(L,p,3p-1)=\sum_{n_{-1},{\bf m}
\atop m_{p-2}\equiv 0~({\rm mod}2)}
 q^{n_{-1}^2-n_{-1}m_0+
{1\over 4}{\bf m{\tilde C}}_{p-1}{\bf
m}}{L-n_{-1}+m_0\atopwithdelims[] n_{-1}}_q\prod_{j=0}^{p-2}
{({1\over 2}{\bf \tilde I}_{p-1}{\bf m}+n_{-1}{\bf
e}_{0})_j\atopwithdelims[] m_j}_q.} 
Next we introduce incidence matrix ${\bf I}'_p$
which is obtained 
from a new   $m,n$ system
\eqn\mnof{\eqalign{n_{-1}+m_{-1}&={1\over 2}(m_{-1}+m_0+L)\cr
                   n_0+m_0&={1\over 2}(-m_{-1}+m_0-m_2+L)\cr
                   n_1+m_1&={1\over 2}m_2\cr
                   n_2+m_2&={1\over 2}(m_1+m_0+m_3)\cr
                   n_j+m_j&={1\over 2}(m_{j-1}+m_{j+1})~~{\rm 
for}~~3\leq j\leq p-3\cr
                   n_{p-2}+m_{p-2}&={1\over 2}m_{p-3}\cr}}
as 
\eqn\inewdef{{\bf m}+{\bf n}={1\over 2}{\bf I}'_p{\bf m}+{L\over
                   2}({\bf e}_0+{\bf e}_{-1}).}
This incidence matrix is shown graphically in Fig. 2.
Then we may rewrite ~\phifop~as 
\eqn\newfof{F_{1,5}(L,p,3p-1)=\sum_{{\bf m}\atop m_{p-2}\equiv 0~({\rm
mod}2)} q^{{1\over 4}(({\bf m}-L{\bf e}_{-1})
{\bf C}'_{p}({\bf m}-L{{\bf e}_{-1}}))}
\prod_{j=-1}^{p-2}{({1\over 2}{\bf I'}_{p}{\bf m}+{L\over 2}({\bf
e}_{-1}+{\bf e}_{0}))_j\atopwithdelims[] m_j}_q}
where ${\bf C}'_{p}=2-{\bf I}'_p.$ 
If we now define 
\eqn\tribfo{B_2(L,p,p')
=\sum_{j=-\infty}^{\infty}
\left(q^{j(pp' j+r_mp'-ps_m)}{L\atopwithdelims() p'j}_2-
q^{(jp+r_m)(jp'+s_m)}{L\atopwithdelims() p'j+s_m}_2\right)}
with $r_m$ and $s_m$ as given in ~\rmsm~
we find from extensive computer studies a polynomial generalization
of the identity~\phiofident~
\eqn\pofident{F_{1,5}(L,p,3p-1)=B_2(L,p,3p-1).}

It now remains to associate the identities~\phiofident~and
\pofident~with a relevant perturbation of the model 
 $M(p,3p-1).$ For this model
the perturbation $\phi_{2,1}$ is not relevant because $2p<p'$ (where
here $p'=3p-1$). On the
other hand the integrable perturbation $\phi_{1,5}$ is relevant for
$2p<p'$ and by generalizing the ``duality'' seen in~\rnewrav-\rrav~between
$\phi_{2,1}$ and $\phi_{1,5}$ we identify the bosonic
polynomial~\tribfo~in general with the $\phi_{1,5}$ perturbation. This
identification is discussed further in sec. 9 where we study
the cases $M(p,2p\pm 1)$ of ~\rrav~and can be tested by computing
the order parameter for the models of Kuniba~\rnineteen.

\newsec{Generalizations}

In the preceding sections we have examined a particularly simple
application of the TABL. However from that work it is apparent that
many generalizations are possible. 

\subsec{The perturbation $\phi_{2,1}$ of $M(p,p')$ for $1<p'/p<3/2.$}

For our first generalization we note that 
instead
of taking as our starting point the polynomial
identities ~\schw~ for the models
$(A_1^{(1)})_2\times (A_1^{(1)})_{p-3}/(A_1^{(1)})_{p-1}$ with $p\geq 4$
an integer as done in sec. 3 we can start with the corresponding
polynomial identities for $p$ fractional which 
are derived in ~\rsw~ from the
identities for the models $M(p,p')$  of \rbm-\rbms.
In sec. 5 of~\rsw~the polynomial identity is proven which generalizes~\schw~
\eqn\costepol{F^{(2)}(L,p,p')=B^{(2)}(L,p,p')~~~{\rm for}~2p'<3p}
where
\eqn\btwo{\eqalign{B^{(2)}(L,p,p')=\sum_{j=-\infty}^{\infty}
&(q^{{j\over 2}[p'(2p-p')j+p'r_0-(2p-p')s_m]}T_0(L,p'j)\cr
&~~~~-q^{{1\over 2}(p'j+s_m)((2p-p')j+r_0)}T_0(L,p'j+s_m))}}
with
\eqn\ntres{r_0=2r_m-s_m,~~~{\rm and}~~|p'r_m-ps_m|=1}
and 
\eqn\ftwo{F^{(2)}(L,p,p')=\sum_{{\bf m}\atop m_{t_f}\equiv ({\rm
mod}2)}q^{{1\over 4}{\bf m}C^{(2)}(p,p'){\bf m}}
\prod_{j=1}^{t_f}{({1\over 2}{\bf I}^{(2)}(p,p'){\bf m}+{L\over 2}{\bf
e}_2)_j\atopwithdelims[] m_j}_q.}
with
\eqn\ct{{\bf C}^{(2)}(p,p')=2-{\bf I}^{(2)}(p,p').}
This identity is completed by specifying the incidence matrix ${\bf
I}^{(2)}(p,p').$ This is done by first introducing the continued
fraction decomposition  
\eqn\confr{{p'\over p'-p}=1+\nu_0+{1\over \nu_1+{1\over \nu_2+\cdots
+{1\over 2+\nu_f}}}}
with $f\geq 1$ and $\nu_0\geq 2$
which defines the numbers $\nu_j.$
From this we define for $f\geq 1$
\eqn\defnf{t_l=\sum_{j=0}^{l}\nu_j,~~~0\leq l \leq f}
and the incidence matrix ${\bf I}^{(2)}(p,p')$ which is given
graphically in Fig. 3
\eqn\twoin{\eqalign{({\bf I}^{(2)}(p,p'))_{1,i}&=\delta_{1,i-1}\cr
({\bf I}^{(2)}(p,p'))_{j,i}&=\delta_{j,i+1}+\delta_{j,i-1}~~~{\rm for}~~
j \neq 1,t_l~~~0\leq l \leq f \cr
({\bf I}^{(2)}(p,p'))_{t_l,i}&=\delta_{t_l,i+1}+\delta_{t_l,i}-
\delta_{t_l,i-1}~~{\rm
for}~~0\leq l \leq f-1\cr
({\bf I}^{(2)}(p,p'))_{t_f,i}&=\delta_{t_f,i+1}+
\delta_{\nu_f,0}\delta_{t_f,i}\cr}}
with $1\leq i,j\leq t_f.$ In the case where $\nu_f=0$  the last 
equation in ~\twoin~is given precedence over the
next to last equation with $l=f-1$.

The TABL may now be applied in the identical manner as for the case $f=0$
which was considered in sec. 3 and the finitization of the 
resulting fermi-bose identity may be made again (and verified in low order
cases on the computer). Thus we find the generalization of ~\polyidto~
to the model $M(p,p')$ with $3p>2p'$
\eqn\otgen{F_{2,1}(L,p,p')=B_1(L,p,p')}
where
\eqn\togen{F_{2,1}(L,p,p')=\sum_{{\bf m}\atop m_{t_f}\equiv 0({\rm mod}2)}
q^{{1\over 4}{\bf m}{\bf \tilde C}^{(2,1)}(p,p'){\bf
m}}\prod_{j=0}^{t_f}{({1\over 2}{\bf \tilde I}^{(2,1)}(p,p'){\bf m}+L{\bf
e}_0)_j\atopwithdelims[] m_j}_q}
where ${\bf \tilde C}^{(2,1)}(p,p')=2-{\bf {\tilde I}}^{(2,1)}(p,p')$
and the incidence matrix shown graphically in Fig. 4 is
\eqn\imatto{\eqalign{
({\bf {\tilde I}}^{(2,1)}(p,p'))_{0,i}&=
-({\bf {\tilde I}}^{(2,1)}(p,p'))_{i,0}=-\delta_{i,2}~{\rm for}~~
0\leq i \leq t_f\cr
({\bf {\tilde I}}^{(2,1)}(p,p'))_{j,i}&=({\bf 
I}^{(2)}(p,p'))_{j,i} ~{\rm for}~~1\leq j,i \leq t_f.\cr}}
We refer to this region where $3p>2p'$
as the region of weak anisotropy for the
$\phi_{2,1}$ perturbation and in this region the result ~\otgen~covers
all cases. In this region the central charge lies in the interval $-2<c<0.$
This is the region studied in~\rlykk~ where there are no vacuum ambiguities.

\subsec{The perturbation $\phi_{1,5}$ for $M(p,4p-p')$
 for $3p>2p'$}

For the second generalization we apply TABL to the trinomial
identity~\otgen~ just as we applied TABL to ~\polyidto~in sec. 5
and thus we obtain for the models $M(p,4p-p')$ with $3p>2p'$ 
the following extension of ~\pofident
\eqn\ofgen{F_{1,5}(L,p,4p-p')
=B_2(L,p,4p-p')}
where
\eqn\fogen{\eqalign{F_{1,5}&(L,p,4p-p')\cr
&=\sum_{{\bf m}\atop m_{t_f}\equiv 0 ({\rm mod}2)}
q^{{1\over 4}(L(L-2m_{-1})+{\bf m}{\bf \tilde C}^{(1,5)}(p,p'){\bf
m})}\prod_{j=-1}^{t_f}{{1\over 2}({\bf \tilde
I}^{(1,5)}(p,p'){\bf m}+L({\bf e}_{-1}+{\bf e}_{0}))_j\atopwithdelims[]
m_j}_q,\cr}}
 ${\bf \tilde C}^{(1,5)}(p,p')=2-{\bf \tilde I}^{(1,5)}(p,p')$ 
and incidence matrix ${\bf \tilde I}^{(1,5)}(p,p')$, 
given graphically in Fig. 5, is
\eqn\inof{\eqalign{({\bf \tilde I}^{(1,5)}(p,p'))_{-1,-1}&=({\bf \tilde
I}^{(1,5)}(p,p'))_{-1,0}= -({\bf \tilde I}^{(1,5)}(p,p'))_{0,-1}=1\cr
({\bf \tilde I}^{(1,5)}(p,p'))_{j,i}
&=({\bf \tilde I}^{(2,1)}(p,p'))_{j,i}+\delta_{j,0}\delta_{i,0}, ~~{\rm
for}~~0\leq j,i\leq t_f\cr}}
and zero otherwise. ${\bf \tilde I}^{(2,1)}(p,p')$ was defined in
~\imatto.

\subsec{Further generalizations for the perturbations 
$\phi_{2,1}$  and $\phi_{5,1}.$ }

We have now generalized the starting point of secs. 3 and 5 and
obtained identities for a wide class of models. 
However since the identity ~\ofgen~is
also in the form~\tbpair~ the TABL may be applied again to obtain a new
identity whose  finitizations can be conjectured and tested on the
computer. As long as the conjectured finitizations are of the form 
~\tbpair~the
process may be continued indefinitely and alternately produces
fermionic representations for
$\phi_{2,1}$ and $\phi_{1,5}$ perturbations. The two new feature of
this procedure is that ``triangles'' appear in the incidence
matrices and there are ``negative tadpoles'' for $\phi_{2,1}$ perturbation. 
The results are as follows: 

{\bf A. The $\phi_{2,1}$ perturbations for $M({\bar p},{\bar p}')$ 
with $3/2<{ \bar p}'/{\bar p}<2,~{\bar p}'\neq 2{\bar p}-2$}

We define ${\bar p}',{\bar p}$  
from $p',p$ of  ~\confr~and an integer $n$ by
\eqn\todonf{{{\bar p}'\over {\bar p}'-{\bar p}}=
2+{1\over n+{1\over {p'\over {p'-p}}-2}},~~
{\rm where}~~n=1,2,3,\cdots ~{\rm and}~{p'\over p'-p}>3,~{p'\over p'-p}\neq 4.}
Next we introduce the $2n+1+t_f$ dimensional incidence matrix 
${\bf J}^{(2,1)}({\bar p},{\bar p}')$ (Fig. 6) as 
\eqn\jmatot{\eqalign{({\bf J}^{(2,1)}({\bar p},{\bar p}'))_{-2n,i}&=
-\delta_{-2n,i}-\delta_{-2n+1,i}+\delta_{-2n+2,i}\cr
({\bf J}^{(2,1)}({\bar p},{\bar p}'))_{i,-2n}&=
-\delta_{-2n,i}+\delta_{-2n+1,i}+\delta_{-2n+2,i}\cr
({\bf J}^{(2,1)}({\bar p},{\bar p}'))_{-2l,i}
&=-\delta_{-2l+1,i}-\delta_{-2l-1,i}+\delta_{-2l-2,i}
+\delta_{-2l+2,i}~{\rm for}~~l=1,2,\cdots n-1\cr
({\bf J}^{(2,1)}({\bar p},{\bar p}'))_{i,-2l}
&=+\delta_{-2l+1,i}+\delta_{-2l-1,i}+\delta_{-2l-2,i}
+\delta_{-2l+2,i}~{\rm for}~~l=1,2,\cdots n-1\cr
({\bf J}^{(2,1)}({\bar p},{\bar p}'))_{-2l-1,i}
&=\delta_{-2l-2,i}+\delta_{-2l-1,i}+\delta_{-2l,i}~{\rm
for}~l=0,1,\cdots ,n-1\cr
({\bf J}^{(2,1)}({\bar p},{\bar p}'))_{i,-2l-1}
&=-\delta_{-2l-2,i}+\delta_{-2l-1,i}-\delta_{-2l,i}~{\rm
for}~l=0,1,\cdots ,n-1\cr
({\bf J}^{(2,1)}({\bar p},{\bar p}'))_{0,i}
&=-\delta_{2,i}+\delta_{0,i}-\delta_{-1,i}+\delta_{-2,i}\cr
({\bf J}^{(2,1)}({\bar p},{\bar p}'))_{i,0}
&=\delta_{2,i}+\delta_{0,i}+\delta_{-1,i}+\delta_{-2,i}\cr}}
for $-2n\leq i \leq t_f$ 
\eqn\newjmatot{({\bf J}^{(2,1)}({\bar p},{\bar p}'))_{j,i}
=({\bf \tilde I}^{(2,1)}(p,p'))_{j,i}~{\rm for}~1\leq j,i \leq  t_f}
and ${\bf C}^{(2,1)}({\bar p},{\bar p}')
=2-{\bf J}^{(2,1)}({\bar p},{\bar p}')$ 
where ${\bf {\tilde I}}^{(2,1)}(p,p')$ is given by \imatto.~
Then defining the
fermionic form
\eqn\pfermtost{F_{2,1}(L,{\bar p},{\bar p}')=\sum_{{\bf m}\atop
m_{t_f}\equiv 0 ({\rm mod}2)}
q^{{1\over 4}{\bf m}{\bf C^{(2,1)}}({\bar p},{\bar p}'){\bf m}}
\prod_{j=-2n}^{t_f}{({1\over 2}{\bf J}^{(2,1)}({\bar p},{\bar p}'){\bf
m}+L{\bf e}_{-2n})_j\atopwithdelims[] m_j}}
and recalling ~\trib~we have the polynomial identity
\eqn\ptostid{F_{2,1}(L,{\bar p},{\bar p}')=
B_1(L,{\bar p},{\bar p}').}
 
{\bf B. The $\phi_{1,5}$ perturbations for $M({\bar p},{\bar p}')$   
with $2<{\bar p}'/{\bar p}<5/2,~{\bar p}'\neq 2{\bar p}+2$}

In the case of the $\phi_{1,5}$ perturbation, generalization is made
in a similar fashion, now
by defining ${\bar p'}$ and ${\bar p}$ as
\eqn\ofdonf{{{\bar p}'\over {\bar p}}=
2+{1\over n+1+{1\over {p'\over p'-p}-2}},~~
{\rm where}~~n=1,2,3,\cdots ~{\rm and}~{p'\over p'-p}>3,
~{p'\over p'-p}\neq 4.}
Then we introduce the $2n+2+t_f$ dimensional incidence matrix 
${\bf J}^{(1,5)}({\bar p},{\bar p}')$ (Fig. 7) as 
\eqn\jofmat{({\bf J}^{(1,5)}({\bar p},{\bar p}'))_{j,i}=
({\bf J}^{(2,1)}({\bar p}'-{\bar p},{\bar p}'))_{j,i}~~-2n-1\leq j,i
\leq t_f}
${\bf C}^{(1,5)}({\bar p},{\bar p}')=
2-{\bf J}^{(1,5)}({\bar p},{\bar p}')$
and the fermionic form
\eqn\strofform{\eqalign{&F_{1,5}(L,{\bar p},{\bar p}')\cr
&=\sum_{{\bf m}\atop m_{t_f}\equiv 0 ({\rm mod}2)}
q^{{1\over 4}(L(L-2m_{-2n-1})+{\bf m}{\bf C}^{(1,5)}({\bar p},{\bar
p}'){\bf
m})}\prod_{j=-1-2n}^{t_f}{{1\over 2}({\bf J}^{(1,5)}({\bar p},{\bar
p}'){\bf m}+
L({\bf e}_{-1-2n}+{\bf e}_{-2n}))_j\atopwithdelims[]
m_j}_q.\cr}}
The  appropriate polynomial identity is
\eqn\stroofid{F_{1,5}(L,{\bar p},{\bar p}')=
B_2(L,{\bar p},{\bar p}').}

\newsec{The model $M(2\nu+1,2(2\nu+1)\pm 2)$ for $\nu \geq 2$}

In this section we consider the important  special case omitted
from the discussion of the previous section: $p'/(p'-p)=4$ in the
continued fractions \todonf~and ~\ofdonf. For this case we
need to find a new starting trinomial identity to replace ~\costepol.
Fortunately such an identity has been derived in~\rbmo~where the
polynomial analogues of the generalized G{\" o}llnitz-Gordon
identities for the $N=1$ supersymmetric model $SM(2,4\nu)$ were derived.
We will show that the direct application of the TABL 
to the trinomial Bailey pair of ~\rbmo~will
lead to the model $M(2\nu+1,4\nu)$ which corresponds to 
the special case $p'/(p'-p)=4$ in ~\todonf.

In ~\rbmo~ it was
proven that with
\eqn\bobmo{B^{N=1}(L,2,4\nu)=\sum_{j=-\infty}^{\infty}(-1)^jq^{\nu
j^2+{j\over 2}}\left(T_0(L,2\nu j)+T_0(L,2\nu j+1)\right),}
the incidence matrix ${\bf I}(2,4\nu)(={\bf I}^{(2)}(2\nu+1,4\nu))$ 
of Fig. 8 defined by
\eqn\inbmo{\eqalign{({\bf I}(2,4\nu))_{1,k}&=\delta_{1,k}-\delta_{2,k}\cr
({\bf I}(2,4\nu))_{j,k}
&=\delta_{j-1,k}+\delta_{j+1,k}~~{\rm for}~~2\leq j
\leq \nu-1\cr
({\bf I}(2,4\nu))_{\nu,k}&=\delta_{\nu-1,k}+\delta_{\nu,k},\cr}}
where $1\leq k \leq \nu,$
the matrix ${\bf C}(2,4\nu)=2-{\bf I}(2,4\nu),$ and the
fermionic form
\eqn\ferbmo{F^{N=1}(L,2,4\nu)=
\sum_{{\bf m}}q^{{1\over 4}(L(L-2m_2)+{\bf m}{\bf
C}(2,4\nu){\bf m})}\prod_{j=1}^{\nu}{{1\over 2}({\bf
I}(2,4\nu){\bf m}+L({\bf e_{1}}+{\bf e}_{2}))_j\atopwithdelims[] m_j}_q}
we have the polynomial identity for $\nu\geq 2$
\eqn\idbmo{F^{N=1}(L,2,4\nu)=B^{N=1}(L,2,4\nu).}
This identity is exactly in the form~\tbpair~ needed for the TABL and thus we
find from ~\ctabl~the character identity for the model $M(2\nu+1,4\nu)$
\eqn\charbmo{\eqalign{\sum_{{\tilde L},{\bf m}}&
q^{{{\tilde L}^2\over 2}+{1\over 4}({\tilde L}({\tilde L}-2m_2)+{\bf
m}{\bf C}(2,4\nu){\bf m})}{1\over (q)_L}\prod_{j=1}^{\nu}{{1\over 2}({\bf
I}(2,4\nu){\bf m}+{\tilde L}({\bf e}_{1}+{\bf e}_{2}))_j\atopwithdelims[] m_j}_q\cr
&=\chi_{\nu+1,2\nu+1}^{(2\nu+1,4\nu)}(q)+
q^{1\over 2}\chi_{\nu,2\nu+1}^{(2\nu+1,4\nu)}(q).\cr}}
Because the right hand side contains both integer and half integer
powers of $q^{{1\over 2}}$ we may split this identity into two
identities for the  characters $\chi_{\nu+1,2\nu+1}^{(2\nu+1,4\nu)}(q)$
and $\chi_{\nu,2\nu+1}^{(2\nu+1,4\nu)}(q)$ separately and we find 
\eqn\spcharid{\eqalign{\sum_{{\tilde L},{\bf m}\atop m_\nu\equiv 
0~({\rm mod}2)}&
q^{{{\tilde L}^2\over 2}+{1\over 4}({\tilde L}({\tilde L}-2m_2)+{\bf
m}{\bf C}(2,4\nu){\bf m})}{1\over (q)_L}\prod_{j=1}^{\nu}{{1\over 2}({\bf
I}(2,4\nu){\bf m}+{\tilde L}({\bf e}_{1}+{\bf e}_{2}))_j\atopwithdelims[] m_j}_q\cr
&=\chi_{\nu+1,2\nu+1}^{(2\nu+1,4\nu)}(q)}}
and similarly for $\chi_{\nu,2\nu+1}^{(2\nu+1,4\nu)}(q)$ with 
$m_{{\nu}}\equiv 1~({\rm mod}2).$ 

We may now finitize this character formula as in previous sections by
defining  ${\bf I}^{(2,1)}(\nu)$ (Fig. 9) as
\eqn\itospe{\eqalign{({\bf I}^{(2,1)}(\nu))_{0,k}&
=-\delta_{0,k}-\delta_{1,k}+\delta_{2,k};~~
({\bf
I}^{(2,1)}(\nu))_{k,0}=-\delta_{0,k}+\delta_{1,k}+\delta_{2,k}~{\rm
for}~0\leq k\leq \nu\cr
({\bf I}^{(2,1)}(\nu))_{j,k}&=
({\bf I}(2,4\nu))_{j,k}~~{\rm for}~~1\leq j,k \leq \nu\cr}}
${\bf C}^{(2,1)}(\nu)=2-{\bf I}^{(2,1)}(\nu)$
 and
write the polynomial identity which has been checked to high order on
the computer 
\eqn\trifrombmo{\sum_{{\bf m}\atop m_{\nu} \equiv 0~({\rm mod}2)}
q^{{1\over 4}{\bf m}{\bf C}^{(2,1)}(\nu){\bf
m}}\prod_{j=0}^{\nu}
{({1\over 2}{\bf I}^{(2,1)}(\nu){\bf m}+L{\bf
e}_0)\atopwithdelims[] m_j}_q=B_1(L,2\nu+1,4\nu)}
Thus we have the polynomial identity for the $\phi_{2,1}$ perturbation
of model $M(2\nu+1,4\nu).$

Since ~\trifrombmo~is of the form ~\tbpair~
we may now apply the TABL once again and make a finitization to
obtain the identity for the $\phi_{1,5}$ perturbation of
$M(2\nu+1,4\nu+4)$
\eqn\spof{\eqalign{&\sum_{{\bf m}\atop m_{\nu} \equiv 0~({\rm mod}2)}
q^{{1\over 4}(L(L-2m_{-1})+{\bf m}{\bf C}^{(1,5)}(\nu){\bf
m})}\prod_{j=-1}^{\nu}
{({1\over 2}{\bf I}^{(1,5)}(\nu){\bf m}+{L\over 2}({\bf
e}_0+{\bf e}_{-1}))\atopwithdelims[] m_j}_q\cr
&=B_2(L,2\nu+1,4\nu+4)\cr}}
where ${\bf I}^{(1,5)}(\nu)$ (Fig. 10) is
\eqn\inofnew{\eqalign{({\bf I}^{(1,5)}(\nu))_{-1,i}
&=\delta_{-1,i}+\delta_{0,i};~~({\bf I}^{(1,5)}(\nu))_{i,-1}
=\delta_{-1,i}-\delta_{0,i}~{\rm for}~-1\leq i \leq \nu \cr
({\bf I}^{(1,5)}(\nu))_{j,i}&=({\bf I}^{(2,1)}(\nu))_{j,i}+
\delta_{j,0}\delta_{i,0}
~~{\rm for }~0\leq i,j\leq \nu\cr}}
with ${\bf I}^{(2,1)}(\nu)$ given by ~\itospe~ 
and
${\bf C}^{(1,5)}(\nu)=2-{\bf I}^{(1,5)}(\nu).$

\newsec{Families of identities}

We now have found identities for the $\phi_{2,1}$ perturbation of
$M(p,2p-2)$ and the $\phi_{1,5}$ perturbation of $M(p,2p+2)$ for all
$p\geq 5$ and thus have found the identities for the special case
$p'/(p'-p)=4$ in~\todonf~and \ofdonf.
However, this is not the end of the story. Just
as in the previous section we could continue to apply the TABL
followed by finitization of the resulting identity to produce an
infinite chain of
identities. Only now the identities will be for models which have been
seen before for which we already had produced an identity and moreover
the new identity we will produce will have a different number of
variables from the old identity. This is new  a phenomena not
previously encountered  which gives
different fermionic representations for the same bosonic polynomial.  
We refer to it as ``families of identities''.

\subsec{The $\phi_{2,1}$ perturbation of $M(2(\nu+n)+1,4(\nu+n))$}

As our first example of such a family we find for the $\phi_{2,1}$
perturbation that the incidence matrix ${\bf I}^{(2,1)}(\nu,n)$
(Fig.11) given by
\eqn\itonn{\eqalign{({\bf I}^{(2,1)}(\nu,n))_{-2n,k}&=-\delta_{-2n,k}-
\delta_{-2n+1,k}+\delta_{-2n+2,k}\cr
({\bf I}^{(2,1)}(\nu,n))_{k,-2n}&=-\delta_{-2n,k}+
\delta_{-2n+1,k}+\delta_{-2n+2,k}\cr
({\bf I}^{(2,1)}(\nu,n))_{-2l,k}&=\delta_{-2l+2,k}-\delta_{-2l+1,k}
-\delta_{-2l-1,k}+\delta_{-2l-2,k}~~{\rm for}~~1\leq l \leq n-1\cr
({\bf I}^{(2,1)}(\nu,n))_{k,-2k}&=\delta_{-2l+2,k}+\delta_{-2l+1,k}
+\delta_{-2l-1,k}+\delta_{-2l-2,k}~~{\rm for}~~1\leq l \leq n-1\cr
({\bf I}^{(2,1)}(\nu,n))_{-2l+1,k}&=\delta_{-2l+2,k}+\delta_{-2l+1,k}
+\delta_{-2l,k}~~{\rm for}~~1\leq l \leq n\cr
({\bf I}^{(2,1)}(\nu,n))_{k,-2l+1}&=-\delta_{-2l+2,k}+\delta_{-2l+1,k}
-\delta_{-2l,k}~~{\rm for}~~1\leq l \leq n\cr
({\bf I}^{(2,1)}(\nu,n))_{0,k}&=-\delta_{1,k}+\delta_{2,k}-
\delta_{-1,k}+\delta_{-2,k}\cr
({\bf I}^{(2,1)}(\nu,n))_{k,0}&=\delta_{1,k}+\delta_{2,k}+
\delta_{-1,k}+\delta_{-2,k}\cr}}
with $-2n \leq k \leq \nu$
\eqn\moreitonn{({\bf I}^{(2,1)}(\nu,n))_{j,k}=({\bf I}(2,4\nu))_{j,k}
~~{\rm for}~~1 \leq k,j\leq \nu}
and ${\bf C}^{(2,1)}(\nu,n)=2-{\bf I}^{(2,1)}(\nu,n)$
give the following trinomial identities for the model $M(2(\nu+n)+1,4(\nu+n))$
\eqn\polytonn{\eqalign{\sum_{{\bf m}\atop m_{\nu}\equiv~0({\rm mod}2)}&
q^{{1\over 4}{\bf m}{\bf C}^{(2,1)}(\nu,n){\bf
m}}\prod_{j=-2n}^{\nu}
{({1\over 2}{\bf I}^{(2,1)}(\nu,n){\bf m}+L{\bf
e}_{-2n})_j\atopwithdelims[] m_j}_q\cr
&=B_1(L,2(\nu+n)+1,4(\nu+n)).}}

The bosonic polynomials in these identities depend on only $\nu+n$
whereas the size of the matrices appearing in the right hand side of
these identities is  $\nu+2n+1$ so that the fermionic 
sides look very different while the bosonic sides are identical. 
Thus for every distinct partition of the number $\nu+n$ into two parts
we obtain different identities.
This is a new phenomenon which we refer to as ``families'' of identities.

\subsec{The perturbation $\phi_{1,5}$ of $M(2(\nu+n)+1,4(\nu+n)+4)$}

Similarly we find for  the $\phi_{(1,5)}$ perturbation that the
incidence matrix ${\bf I}^{(1,5)}(\nu,n)$ (Fig. 12) given by
\eqn\iofnn{({\bf I}^{(1,5)}(\nu,n))_{j,k}
=({\bf I}^{(2,1)}(\nu,n+1))_{j,k}~~{\rm for}~~-2n-1 \leq j,k
\leq \nu}
gives the polynomial identities for the model $M(2(\nu+n)+1,4(\nu+n)+4)$
\eqn\polyofnn{\eqalign{\sum_{{\bf m}\atop m_{\nu}\equiv~0({\rm mod}2)}&
q^{{1\over 4}(L(L-m_{-2n-1})+{\bf m}{\bf C}^{(1,5)}(\nu,n){\bf
m})}\prod_{j=-2n-1}^{\nu}
{({1\over 2}{\bf I}^{(1,5)}(\nu,n){\bf m}+{L\over 2}({\bf
e}_{-2n}+{\bf e}_{-2n-1}))_j\atopwithdelims[] m_j}_q\cr
&=B_2(L,2(\nu+n)+1,4(\nu+n)+4)}}
with ${\bf C}^{(1,5)}(\nu,n)=2-{\bf I}^{(1,5)}(\nu,n).$
\subsec{The family of identities for the $\phi_{2,1}$ perturbation
of $M({\bar p},{\bar p}')$ for $3/2<{\bar p}'/{\bar p}<2.$}

The phenomena of families of fermionic representations of the
same bosonic characters is not limited to the models $M(p,2p\pm 2).$
To see this we further generalize the starting identity ~\idbmo. We do this
by noting that the identity~\costepol~ which is valid when in
the decomposition ~\confr~$\nu_0\geq 2$~may be extended to the case
$\nu_0=1$ by defining for this case the incidence matrix 
${\bf I}^{(2)'}(p,p')$ (Fig. 13) as
\eqn\instr{\eqalign{({\bf I}^{(2)'}(p,p'))_{j,i}&
=\delta_{j-1,i}+\delta_{j+1,i}~{\rm
for}~~j \neq t'_l,~~1\leq l \leq f\cr
({\bf I}^{(2)'}(p,p'))_{1,i}&=\delta_{1,i}-\delta_{2,i}\cr
({\bf I}^{(2)'}(p,p'))_{t'_l,i}&=\delta_{-1+t'_{l},i}+\delta_{t'_l,i}
-\delta_{1+t'_l,i}~{\rm
for}~1\leq l \leq f-1\cr
({\bf I}^{(2)'}(p,p'))_{t'_f,i}&=\delta_{-1+t'_f,i}+\delta_{\nu_f,0}
\delta_{{t'_f},i}\cr}}
with ~~$1\leq i,j \leq t'_f,~~t'_l=1+\sum_{j=1}^{l}\nu_j$
and the fermionic form $F^{(2)'}(L,p,p')$  as
\eqn\strferp{F^{(2)'}(L,p,p')=\sum_{{\bf m}\atop m_{t_f}\equiv 0~({\rm
mod}2)}q^{{1\over
4}(L(L-2m_2)+{\bf m}{\bf C}^{(2)'}(p,p'){\bf m})}\prod_{j=1}^{t'_f}{({1\over
2}{\bf I}^{(2)'}(p,p'){\bf m}+{L\over 2}
({\bf e}_1+{\bf e}_2))_j \atopwithdelims[] m_j}_q}
with ${\bf C}^{(2)'}(p,p')=2-{\bf I}^{(2)'}(p,p')$
to obtain the identity
\eqn\stridn{F^{(2)'}(L,p,p')=B^{(2)}(L,p,p')}
with $B^{(2)}(L,p,p')$ given by ~\btwo~and
\eqn\ynew{{p'\over p'-p}=2+{1\over \nu_1+
{1\over \nu_2+\cdots {1\over 2+\nu_f}}}}
Special cases of ~\stridn~ are easily verified on the
computer and a general proof will be given elsewhere.

Using this as a starting identity we proceed as before and for $f\geq 2$ 
define  ${\bar p}', {\bar p}$ from $p,p'$ and an integer
$n\geq 0$ as  
\eqn\newpk{{{\bar p}'\over {\bar p}'-{\bar p}}=2+{1\over n+\nu_1+{1\over
 \nu_2+\cdots{1\over \nu_f+2}}}}
We then obtain the identities
\eqn\newtoind{\eqalign{\sum_{{\bf m}\atop m_{t_f}\equiv~0({\rm mod}2)}&
q^{{1\over 4}{\bf m}{\bf C}^{(2,1)}(p,p',n){\bf
m}}\prod_{j=-2n}^{t'_f}
{({1\over 2}{\bf I}^{(2,1)}(p,p',n){\bf m}+L{\bf
e}_{-2n})_j\atopwithdelims[] m_j}_q\cr
&=B^{(2,1)}(L,{\bar p},{\bar p}')}}
where ${\bf C}^{(2,1)}(p,p',n)=2-{\bf J}^{2,1)}(p,p',n)$
and
 ${\bf I}^{(2,1)}(p,p',n),$ given graphically by Fig.14, is
\eqn\newinto{\eqalign{({\bf I}^{(2,1)}(p,p',n))_{i,j}&=({\bf
I}^{(2,1)}(\nu,n))_{i,j}~~-2n\leq i,j \leq 2\cr
({\bf I}^{(2,1)}(p,p',n))_{j,i}&=({\bf
I}^{(2)'}(p,p'))_{j,i}~~1\leq j,i\leq t'_f}}
and zero otherwise.
Clearly ~\newtoind~reduces to ~\polytonn~when $f=2, \nu_2=0$ and $\nu=\nu_1+1.$
We also note that the models described by these representations have
already been studied in sec. 6 in terms of the incidence matrix ${\bf
J}^{(2,1)}({\bar p},{\bar p}').$

When $f=1$ we replace ~\newpk~by
\eqn\fonedef{{{\bar p}'\over {\bar p}'-{\bar p}}=2+{1\over n+2+\nu_1}}
and with this definition the results \newtoind~and \newinto~are
unchanged. This case is the model $M({\bar p},2{\bar p}-1)$ with
${\bar p}=3+n+\nu_1.$

\subsec{The family of identities for the $\phi_{1,5}$ perturbation
of $M({\bar p},{\bar p}')$ for $2<{\bar p}'/{\bar p}<5/2.$}

Similarly for the perturbations $\phi_{1,5}$ we define for $f\geq 2$
another pair ${\bar p},{\bar p}'$ from the pair $p,p'$ as given in ~\ynew~
and an integer $n\geq 0$ as
\eqn\another{{{\bar p}'\over {\bar p}}=2+{1\over n+1+\nu_1+{1\over
\nu_2+\cdots +{1\over \nu_f+2}}}}
and find the family of identities
\eqn\newofind{\eqalign{\sum_{{\bf m}\atop m_{t'_f}\equiv~0({\rm mod}2)}&
q^{{1\over 4}(L(L-2m_{-2n-1})+{\bf m}{\bf C}^{(1,5)}(p,p',n){\bf
m})}\prod_{j=-2n-1}^{t'_f}
{{1\over 2}({\bf I}^{(1,5)}(p,p',n){\bf m}+L({\bf
e}_{-2n}+{\bf e}_{-2n-1}))_j\atopwithdelims[] m_j}_q\cr
&=B_1(L,{\bar p},{\bar p}')\cr}}
where ${\bf C}^{(1,5)}(p,p',n)=2-{\bf I}^{(1,5)}(p,p',n)$ and 
${\bf I}^{(1,5)}(p,p',n),$ given graphically in fig.15 is
\eqn\newinof{({\bf I}^{(1,5)}(p,p',n))_{i,j}=({\bf
I}^{(2,1)}(p'-p,p',n+1))_{i,j}~~-2n-1\leq i,j \leq t'_f}
which reduces to~\polyofnn~when $f=2, \nu_2=0$ and $\nu=\nu_1+1.$

When $f=1$ \another~ is replaced by
\eqn\anothergain{{{\bar p}'\over{\bar p}}=2+{1\over 3+n+\nu_1}~~{\rm
with}~n\geq 0,\nu_{1}\geq 1}
and with this definition ~\newofind~and~\newinof~continue to hold. In
this case the model is $M({\bar p},2{\bar p}+1)$ with ${\bar p}=3+n+\nu_1.$ 

\newsec{Additional fermionic representations for the models $M(p,2p\pm1)$}

For the models $M(2p,2p\pm 1)$ there is yet one further representation
which has not been obtained in sec. 8. This representation is needed
in order to make contact with ~\rrav. To derive this additional representation
we use an identity proven by Andrews~\rnewand~for
 the first model of the series $M(2,5)$ 
\eqn\rrtri{\sum_{n=0}^{\infty}q^{n^2}{L-n\atopwithdelims[]
n}_q=B_2(L,2,5)}
where $B_2(L,2,5)$ is defined by~\tribfo.
An alternative proof is given in the appendix.
We note that the left hand side of ~\rrtri~is the same as that of the
polynomial analogue of  the Rogers-Ramanujan identities 
~\rmac-\randrews.  The right hand side is however of a
very different form than~\rschur~ because it involves  
$q$-trinomials instead of
$q$-binomials.

We can now apply the TABL~\ctabl~ to ~\rrtri~and (letting $(L,n)\rightarrow
(m_1,m_2$))we obtain
\eqn\arrav{\sum_{m_1,m_2=0}^{\infty}q^{m^2_2+m^2_1-m_1m_2}{1\over
(q)_{m_1}}{m_1-m_2\atopwithdelims[] m_2}_q=\chi^{(3,5)}_{1,2}.}
This can be generalized to the polynomial identity
\eqn\brrav{
\sum_{m_1,m_2=0}^{\infty}q^{m_1^2+m_2^2-m_1m_2}{L-m_1+m_2\atopwithdelims[]
m_1}_q{m_1-m_2\atopwithdelims[] m_2}_q=B_1(L,3,5)}
which has been checked to high orders on the computer.
The fermionic side of ~\brrav~is of the canonical form~\fermiform~where
${\bf B}={\bf C}_{2}$ is the Cartan matrix for $A_2$ as defined in ~\in.

This procedure may now be repeated indefinitely as done in the
previous sections and we obtain the following trinomial identities:
\eqn\crrav{\sum_{\bf m} q^{{1\over2}{\bf m}{\bf C}_n{\bf
m}}\prod_{j=1}^{n}{(1-{\bf C}_n){\bf m}+L{\bf e}_1)_j\atopwithdelims[]
m_j}_q=\cases{B_1(L,{n+4\over 2},n+3)&for $n$ even\cr
B_2(L,{n+3\over 2},n+4)&for $n$ odd.\cr}}
where ${\bf C}_n$  is defined in ~\in.
The matrix in the quadratic form  ${\bf C}_n$ is
exactly the same as the TBA matrix $T_1\times A_{n}$ proposed in
~\rrav~ for the $\phi_{2,1}$ perturbation of $M({n+4\over 2},n+3)$ for $n$
even and for the $\phi_{1,5}$ perturbation of $M({n+3\over 2},n+4)$
for $n$ odd. This confirms our previous identification of the
polynomial $B_1(L,p,p')$ with the $\phi_{2,1}$
perturbation and provides the most direct evidence that
$B_2(L,p,p')$ is to be identified with the $\phi_{1,5}$
perturbation.

\newsec{Discussion}

We have now completed our study of the use of the trinomial analogue
of Bailey's lemma to give the Rogers-Ramanujan type identities
relevant for  the $\phi_{2,1}$ and $\phi_{1,5}$
perturbations of conformal field theory. However, before we conclude
we wish to make several observations about our work.

First, it should be remarked, that we have presented here many
conjectured finitizations, which, although amply verified on the
computer, remain to be proven. In principle this is can be done by
using  $L$ difference equations, telescopic expansions, and the
properties of q-trinomials. This method, however usually involves the
use of many of the characters for which we have not yet written down
formula and as seen in ~\rbms~some of these character formulas are
cumbersome once  the
number $f$ in the continued fraction expansions like \confr~ becomes
large. For that reason it would be useful to extend to $\phi_{2,1}$
and $\phi_{1,5}$ perturbations the methods of
Burge~\rburge~which have been recently employed~\romar~ to  study
$\phi_{1,3}$
representations of some of the $M(p,p')$ characters. This method has
the virtue of allowing the study of the characters with the minimal
conformal dimensions $\chi_{r_m,s_m}^{(p,p')}(q)$ in
isolation from the other characters.

More important than the explicit proofs,  however, are
the implications and interpretations of our results in terms of 
perturbed conformal
field theory and its relation to lattice statistical mechanics. In
particular we will here discuss the phenomena of families of identities
found in sec. 8.

In sec. 8 we found for the perturbation  $\phi_{2,1}$ of $M(p,p')$
with $3p<2p'$ and for the perturbation $\phi_{1,5}$ of $M(p,p')$ with 
$2<p'/p < 5/2$ that we had not just one but a family of
fermionic representation for the same bosonic polynomial. These
different fermionic representations should give differerent TBA
systems and the relation of this multiplicity of representations to
conformal field theory needs to be elucidated.

It seems to us to be significant that for the $\phi_{2,1}$ perturbations
the fermionic representations found in sec. 4 were unique when
$3p>2p'$ and that this is exactly the region where perturbative
studies~\rlykk~have no problems with vacuum identification. 
The non uniqueness we found in section 8 sets in at exactly the
regimes where the study of~\rlykk~encountered problems in
identification of the vacuum.
This is a new
phenomena in the study of massive deformations of conformal field
theory and needs much further investigation.

One of the most promising ways to make such a study is to extend the
study of the dilute $A_n$ models to fractional levels and to obtain
the Bethe Ansatz equations for the systems. In these equations a
change of vacuum would be signaled if there were a multiplicity of
regions of the spectral variable such that in the differerent regions
differerent string configurations of the roots which compose the
vacuum were allowed.

However this is not the only possibility of interpreting our
results. It has been known for many years
that in the
``repulsive'' regime  the sine Gordon/massive Thirring model has 
several differerent lattice regularizations which lead to differerent
renormalized theories with differerent properties. One of these
regularizations is based on the XYZ model~\rjkm-\rjnw~and the other
is based on the R matrix of the XXZ model~\rkor-\rbk.
These different
regularizations represent different ways to fill the Fermi sea and can lead
to different physical results because in the ``repulsive'' regimes of
the sine Gordon theory there is attraction between the particles in the
Fermi sea. Such a mechanism
also involves a vacuum choice in field theory and could be relevant to 
the interpretation of our results. 

\appendix{A} {Proof of 9.1}
To prove ~\rrtri~we   send $q\rightarrow
1/q,$ multiply by $q^{L^2\over 2}$,  use \qbininv, 
 define $m$ by $m+n=(m+L)/2$ so that $L-n=(m+L)/2,$
and use the definition ~\qtri~with $n=0$ to write it as
\eqn\appa{\sum_{m=0}^{\infty}q^{{L^2\over 4}+{m^2\over 4}}{{1\over
2}(L+m)\atopwithdelims[] m}_q=\sum_{j=-\infty}^{\infty}
q^{{5\over 2}j^2+j}[T_0(L,5j)-T_0(L,5j+2)].}
We call the series on the left hand side $\phi(L)$ and use the
recursion relations for q-binomials 
\eqn\appb{{L \atopwithdelims[] m}_q={L-1\atopwithdelims[]
m-1}_q+q^m{L-1\atopwithdelims[] m}_q=q^{L-m}{
L-1\atopwithdelims[]m-1}_q+{L-1\atopwithdelims[] m}_q}
to show that
\eqn\appc{\phi(L)=q^{L-{1\over 2}}\phi(L-1)+q^{L-1}\phi(L-2).}
We will prove that \appa~ holds by proving that the right hand side of
~\appa~ which we denote as $B(L)$ satisfies ~\appb~ and agrees with
$\phi(L)$ for $L=0,1.$

From the definition of $B(L)$ we see that
\eqn\appd{\eqalign{&B(L)-q^{L-{1\over
2}}B(L-1)-q^{L-1}B(L-2)\cr
&=\sum_{j=-\infty}^{\infty}q^{{5\over
2}j^2+j}[T_0(L,5j)-q^{L-1/2}T_0(L-1,5j)-q^{L-1}T_0(L-2,5j)\cr
&~~~~-T_0(L,5j+2)+q^{L-1/2}T_0(L-1,5j+2)+q^{L-1}T_0(L-2,5j+2)].\cr}}
This is simplified first by using (4.1) of \rbmo~
\eqn\appe{\eqalign{&T_0(L,A)-q^{L-1/2}T_0(L-1,A)-q^{L-1}T_0(L-2,A)\cr
&~~~=T_0(L-1,A+1)+T_0(L-1,A-1)-T_0(L-2,A)}}
to reduce the right hand side of~\appd~ to
\eqn\appf{\sum_{j=-\infty}^{\infty}
q^{{5\over
2}j^2+j}[T_0(L-1,5j-1)-T_0(L-1,5j+3)-T_0(L-2,5j)+T_0(L-2,5j+2)].}
The second and fourth terms are combined by use of (4.3) of \rbmo~
\eqn\appg{T_0(L,A)-T_0(L-1,A-1)=q^{A+1/2}[T_0(L,A+1)-T_0(L-1,A+2)]}
with $A=5j+3$ to obtain
\eqn\apph{\eqalign{\sum_{j=-\infty}^{\infty}&
(q^{{5\over 2}j^2+j}[T_0(L-1,5j-1)-T_0(L-2,5j)]\cr
&-q^{{{5\over 2}j^2}+6j+
{7\over 2}}[T_0(L-1,5j+4)-T_0(L-2,5j+5)])\cr}}
which is seen to vanish if in the last two terms we let $j\rightarrow j-1$
and hence $B(L)$ satisfies the equation for $\phi(L)$~\appc. It is
easily verified that $\phi(L)=B(L)$ for $L=1,2$ and thus we have
proven~\appa~and~\rrtri~as desired.
\bigskip
{\bf Acknowledgments}

We are pleased to acknowledge useful conversations with G.E. Andrews,
 V.V. Bazhanov, R.M. Ellem, P. Fendley, V. Korepin, S.O. Warnaar, and
A. Schilling.
We also wish to thank M. Batchelor, R.J. Baxter and V.V. Bazhanov for
hospitality at the Australian National University where this work was started.
This work is supported in part by the NSF under DMR9703543 and by the
Australian Research Council.

\vfill
\eject

{\bf Figures}

\bigskip 
\centerline{\epsfxsize=6in\epsfbox{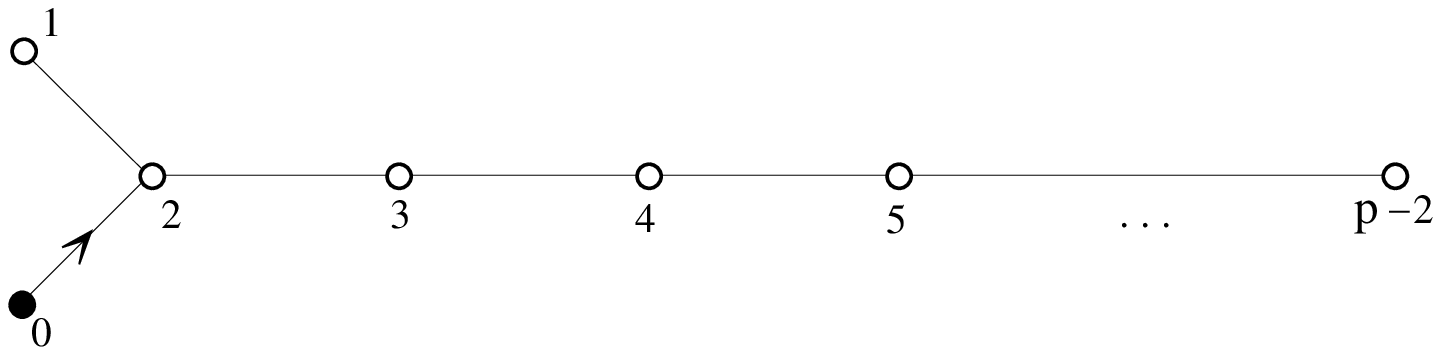}}
\nobreak
Fig.1 Incidence matrix ${\tilde {\bf I}}_{p-1}$~\idefn~ for the
$\phi_{2,1}$ perturbation of $M(p,p+1).$
We use the convention that if sites
$a$ and $b$ are connected by a line then 
$({\bf \tilde I}_{p-1})_{a,b}=({\bf \tilde I}_{p-1})_{b,a}=1$ and if
$a$ and $b$ are joined by an arrow pointing from $a$ to $b$ then
$({\bf\tilde I}_{p-1})_{a,b}=-({\bf \tilde I}_{p-1})_{b,a}=-1.$ 
If a site $a$ is  specified by a filled circle
there is an inhomogeneous term $L{\bf e}_a$ in the $m,n$ system~\nmsy. 
These conventions will be used in all the diagrams of this paper.

\bigskip
\centerline{\epsfxsize=6in\epsfbox{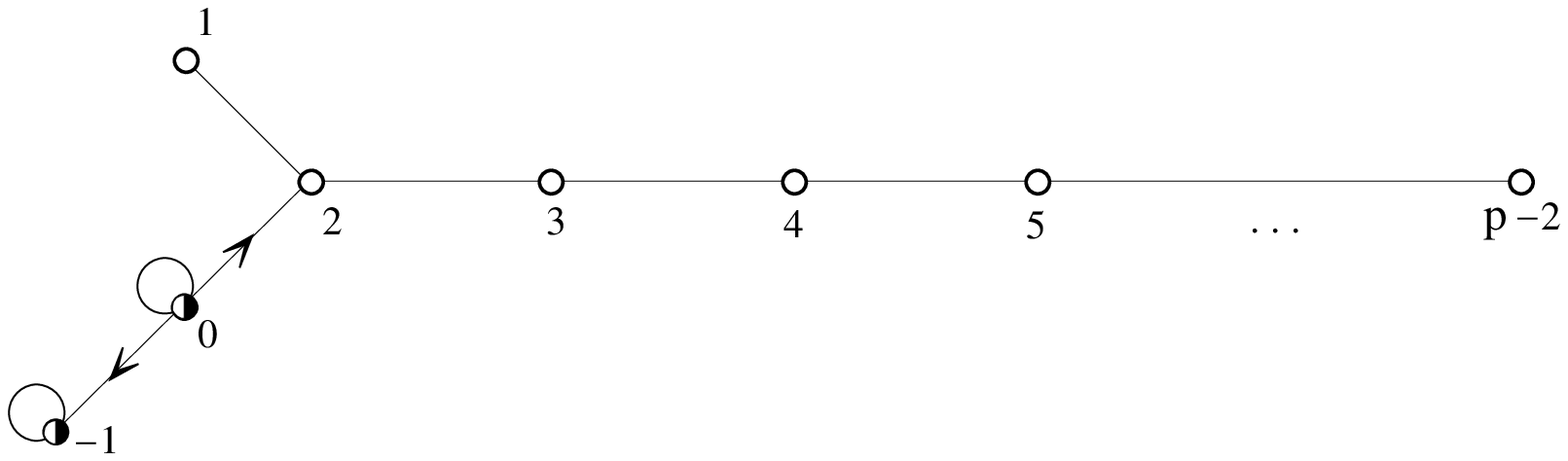}}
\nobreak
Fig.2 Incidence matrix ${\bf I}'_{p}$ appearing in~\inewdef~ for the $\phi_{1,5}$
perturbation of the model $M(p,3p-1).$ The ``tadpole'' at
site $a$ indicates that $({\bf I}'_{p})_{a,a}=1.$ The half filled
circle at site $a$ indicates an inhomogeneous term ${L\over 2}{\bf
e}_a$ in the $m,n$ system \inewdef. These conventions will be used throughout
the paper.

\bigskip
\centerline{\epsfxsize=6in\epsfbox{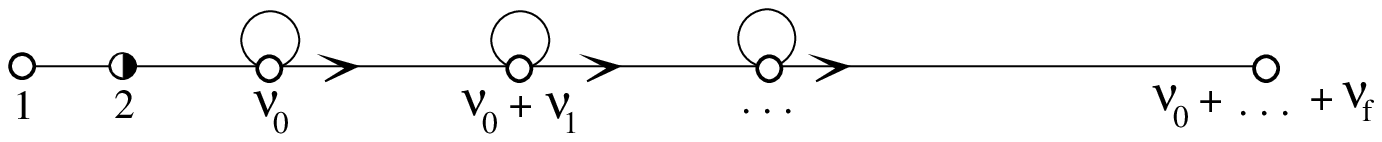}}
\nobreak
Fig.3 Incidence matrix ${\bf I}^{(2)}(p,p')$~\twoin. 
The relation between $p,p'$ and $\nu_0,\cdots ,\nu_f$ is given in ~\confr
The sites
between the tadpole sites are not explicitly shown. 
The arrow following a tadpole at site $a$ connects to the
 site $a+1.$ The remaining sites are connected by lines (not
 arrows). This convention is used in all subsequent diagrams. 

\bigskip
\centerline{\epsfxsize=6in\epsfbox{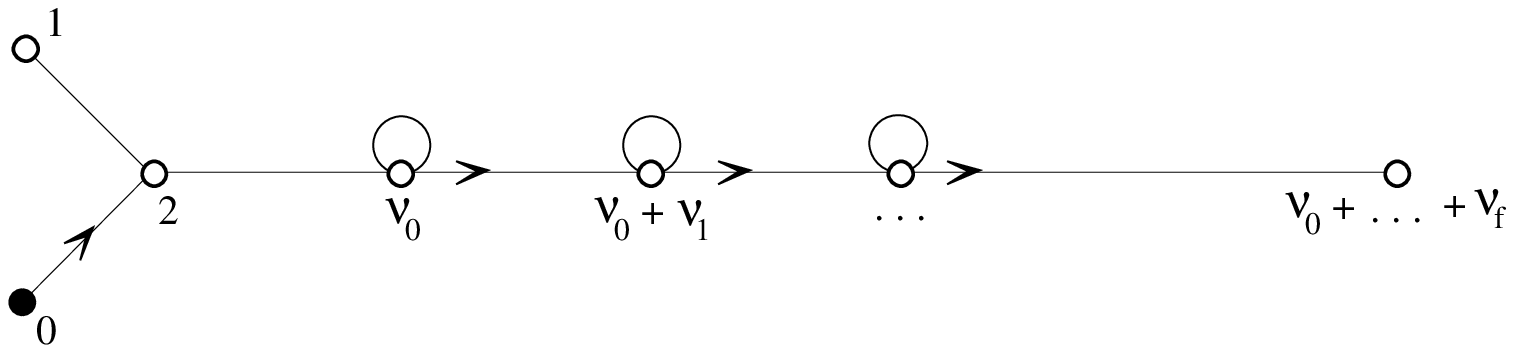}}
\nobreak
Fig.4 Incidence matrix ${\bf \tilde I}^{(2,1)}(p,p')$~\imatto~
for the $\phi_{2,1}$ perturbation of $M(p,p')$ with $1<p'/p<3/2.$ The
relation between $p,p'$ and $\nu_0,\cdots , \nu_f$ is given in ~\confr.
\bigskip
\centerline{\epsfxsize=6in\epsfbox{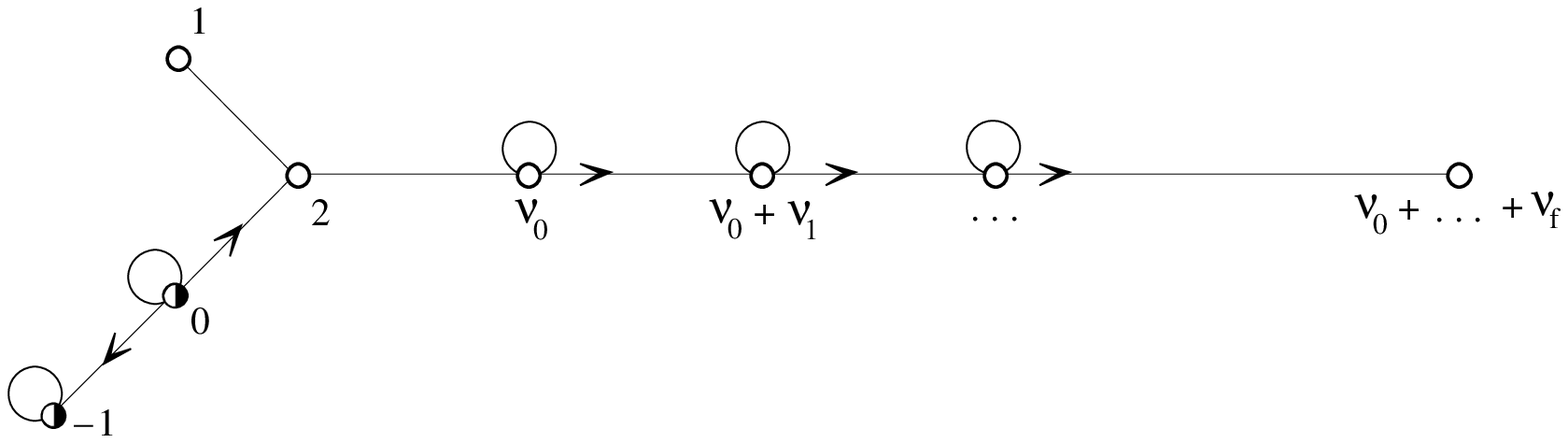}}
\nobreak
Fig.5 Incidence matrix ${\bf \tilde I}^{(1,5)}(p,p')$~\inof~ for the
$\phi_{1,5}$ perturbation of $M(p,4p-p')$ with $1<p'/p<3/2.$
The
relation between $p,p'$ and $\nu_0,\cdots , \nu_f$ is given in ~\confr.
\bigskip
\centerline{\epsfxsize=6in\epsfbox{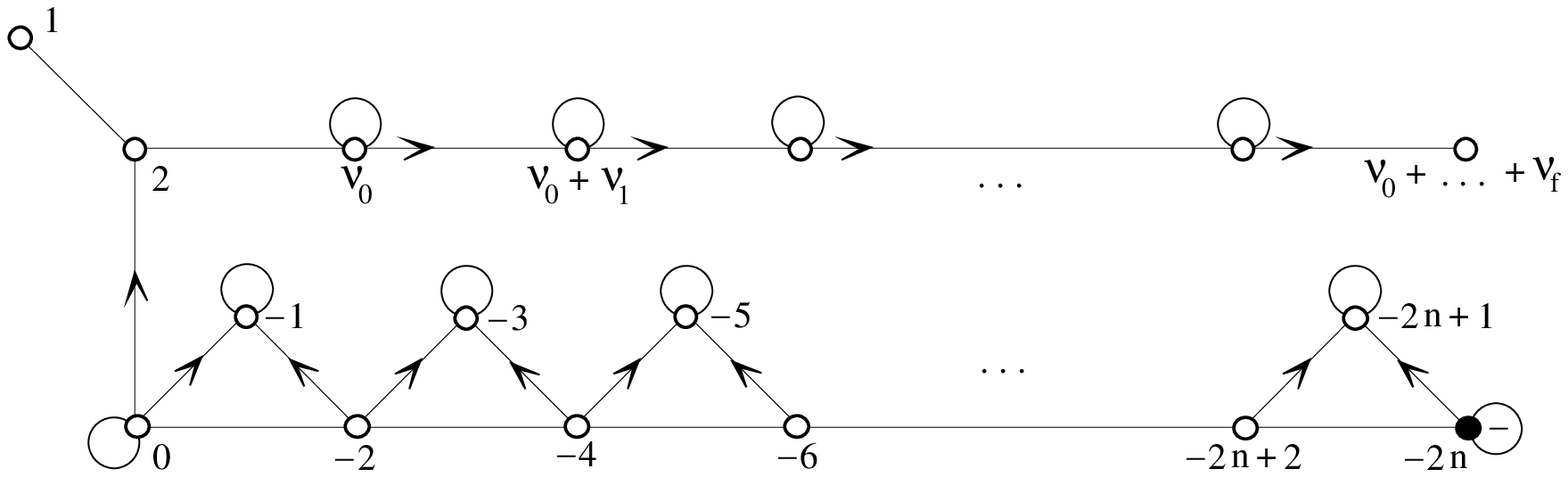}}
\nobreak
Fig.6 Incidence matrix ${\bf J}^{(2,1)}({\bar p},{\bar p}')$ ~\jmatot~for
$\phi_{2,1}$ perturbations of $M({\bar p},{\bar p}')$ with 
$3/2<{\bar p}'/{\bar p}<2$ and
${\bar p}'\neq  2{\bar p}-2.$ 
The relation between ${\bar p},{\bar p}'$ and $\nu _0,\cdots \nu_f,n$ 
is given in ~\confr,\todonf.
The minus
sign inside the tadpole at site $-2n$ indicates that
$({\bf J}^{(2,1)}({\bar p},{\bar p}'))_{-2n,-2n}=-1$.  
This convention for the ``negative tadpole'' will be
used in all subsequent diagrams.

\bigskip
\centerline{\epsfxsize=6in\epsfbox{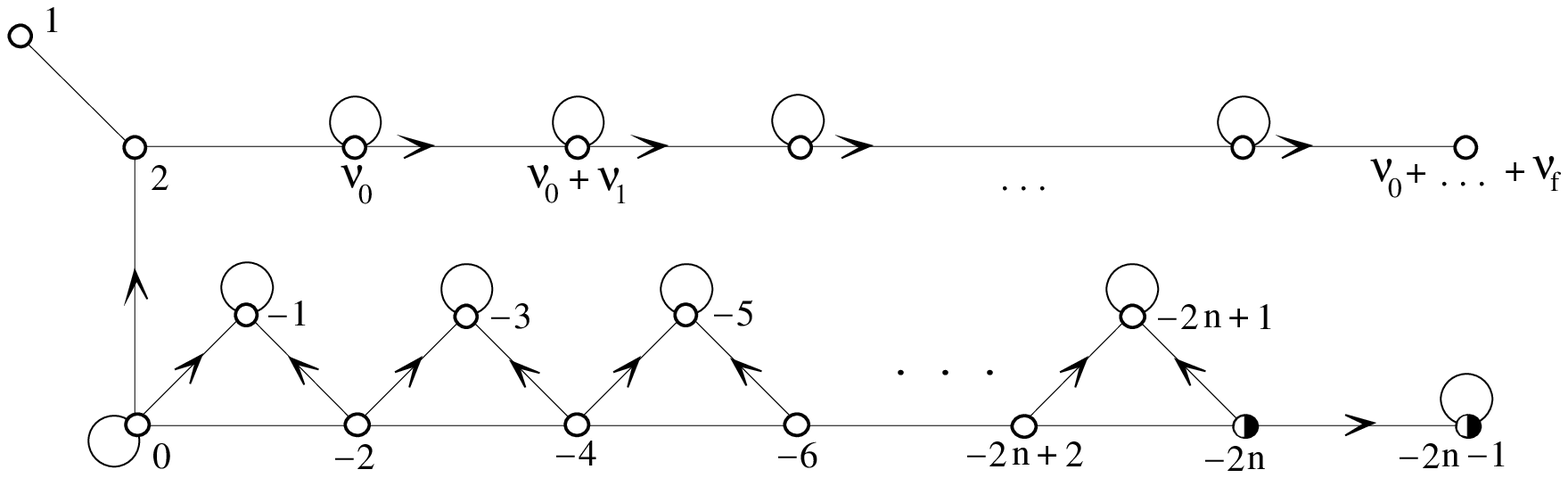}}
\nobreak
Fig.7 Incidence matrix ${\bf J}^{(1,5)}({\bar p},{\bar p}')$~\jofmat~ for
$\phi_{1,5}$ perturbations of $M({\bar p},{\bar p}')$ 
with $2<{\bar p}'/{\bar p}<5/2$ and ${\bar p}'\neq 2{\bar p}+2.$ 
The relation between ${\bar p},{\bar
p}'$ and $\nu_0, \cdots, \nu_f,n$ is given in ~\ofdonf.

\bigskip
\centerline{\epsfxsize=6in\epsfbox{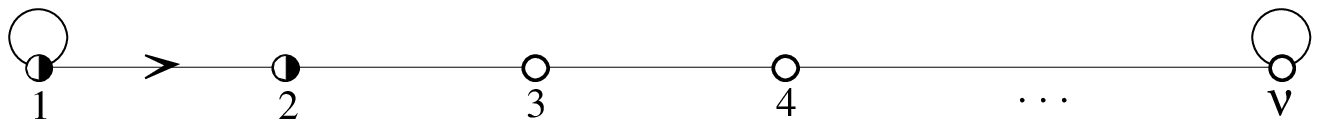}}
\nobreak
Fig.8 Incidence matrix  ${\bf I}(2,4\nu)$~\inbmo~ for the
characters of the $N=1$ supersymmetric model $SM(2,4\nu).$

\bigskip
\centerline{\epsfbox{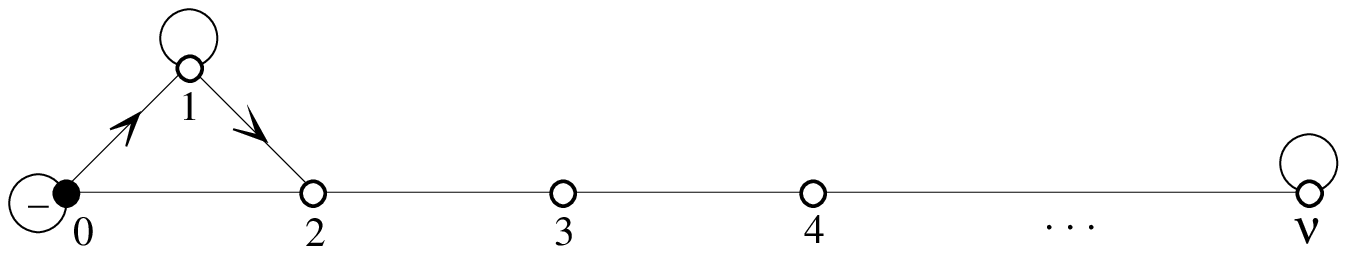}}
\nobreak
Fig.9 Incidence matrix ${\bf I}^{(2,1)}(\nu)$~\itospe~ for the
$\phi_{2,1}$ perturbation of $M(2\nu+1,4\nu)$

 \centerline{\epsfxsize=6in\epsfbox{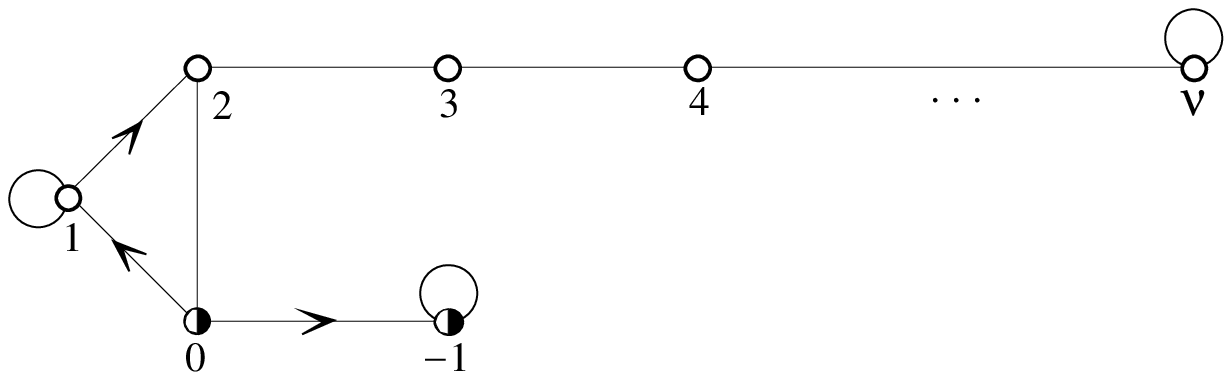}}
Fig.10 Incidence matrix ${\bf I}^{(1,5)}(\nu)$~\inofnew~ for the
$\phi_{1,5}$ perturbation of $M(2\nu+1,4\nu+4)$

\bigskip
 \centerline{\epsfxsize=6in\epsfbox{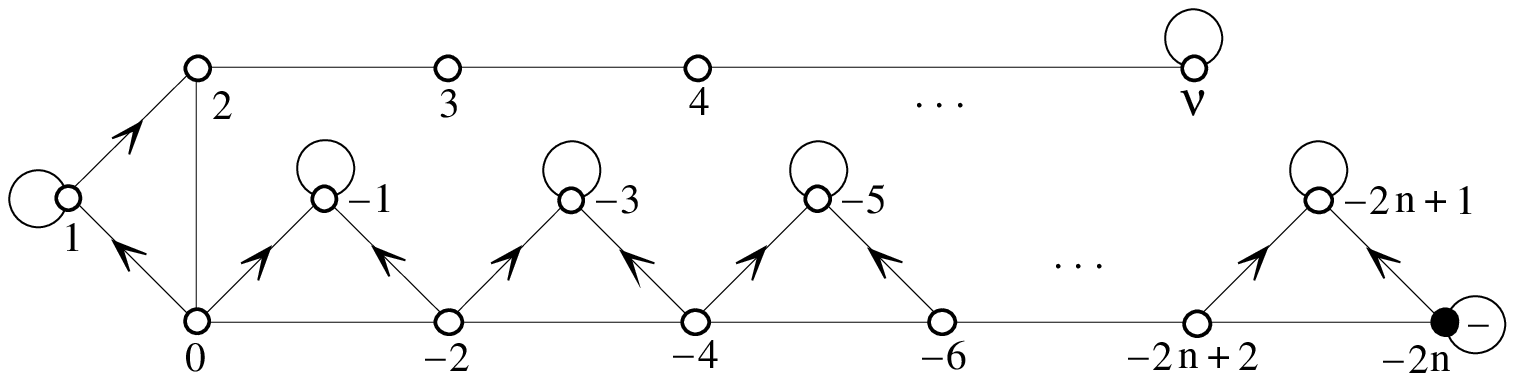}}
\nobreak
Fig.11 Incidence matrix ${\bf I}^{(2,1)}(\nu,n)$~\itonn~ for the
families of identities for $\phi_{2,1}$ perturbations of 
$M(2(\nu+n)+1,4(\nu+n)).$

\bigskip
 \centerline{\epsfxsize=6in\epsfbox{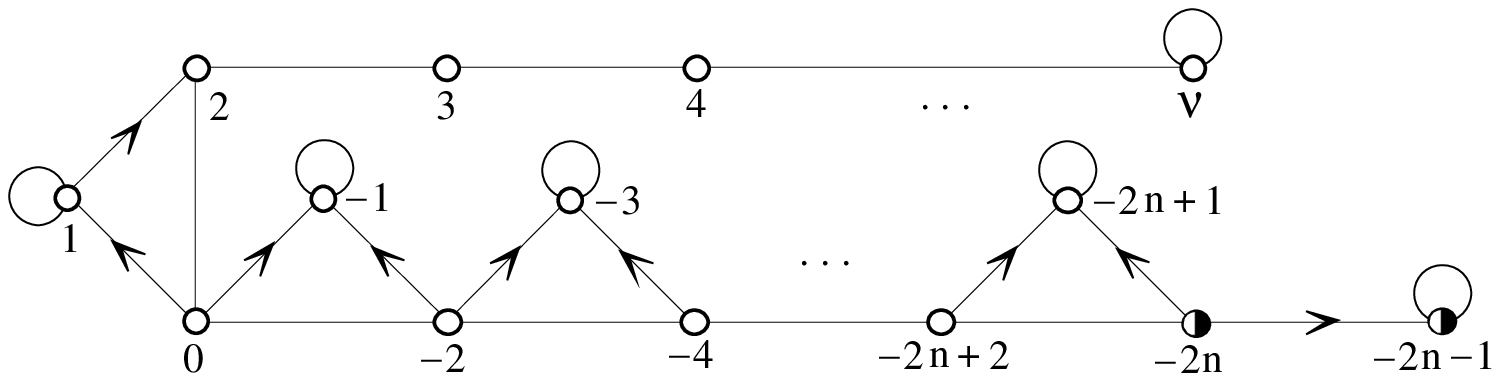}}
\nobreak
Fig.12 Incidence matrix ${\bf I}^{(1,5)}(\nu,n)$~\iofnn~ for the
family of identities for $\phi_{1,5}$ perturbations of
$M(2(\nu+n)+1,4(\nu+n)+4).$

\bigskip
 \centerline{\epsfxsize=6in\epsfbox{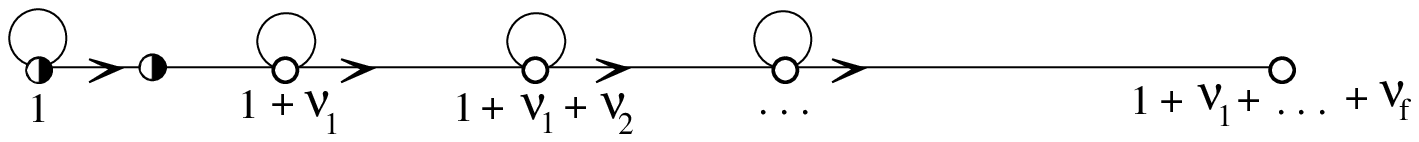}}
\nobreak
Fig.13 Incidence matrix ${\bf I}^{(2)'}(p,p')$~\instr~. Note that for
$f=1$ the only tadpole is at site 1.The relation between $p,p'$ and
$\nu_1, \cdots, \nu_f$ is given in ~\ynew.

\bigskip
 \centerline{\epsfxsize=6in\epsfbox{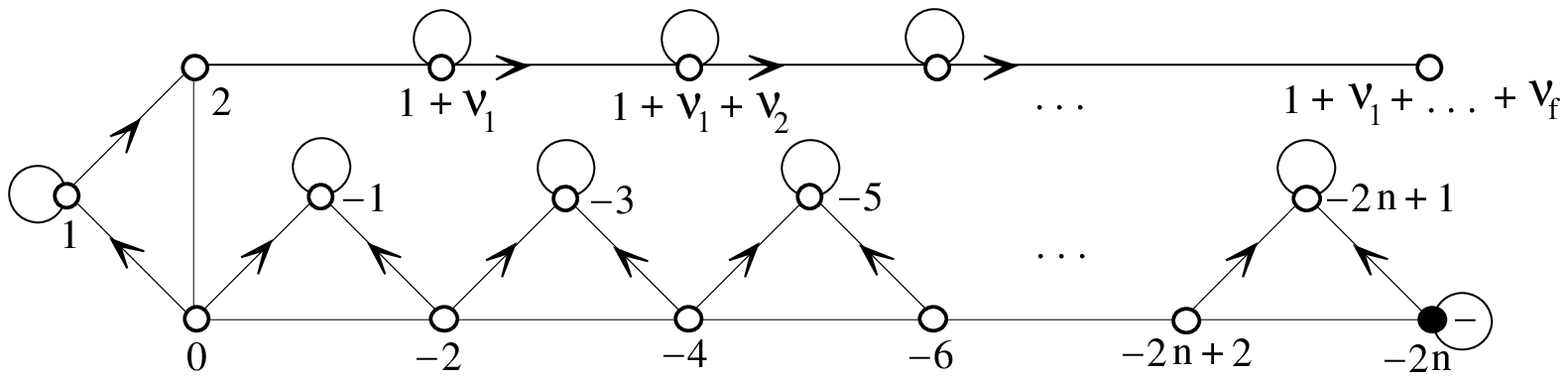}}
\nobreak
Fig.14 Incidence matrix ${\bf I}^{(2,1)}(p,p',n)$~\newinto~ for
the $\phi_{2,1}$ perturbation family of $M({\bar p},{\bar p}')$ 
for $3/2<{\bar p}'/{\bar p}<2.$ The relation between ${\bar p},~{\bar p}'$ and
$p,p',n$ is given in \ynew, \newpk.
If $f=2$ and $\nu_2=0$ this diagram reduces to Fig. 11.
\bigskip
 \centerline{\epsfxsize=6in\epsfbox{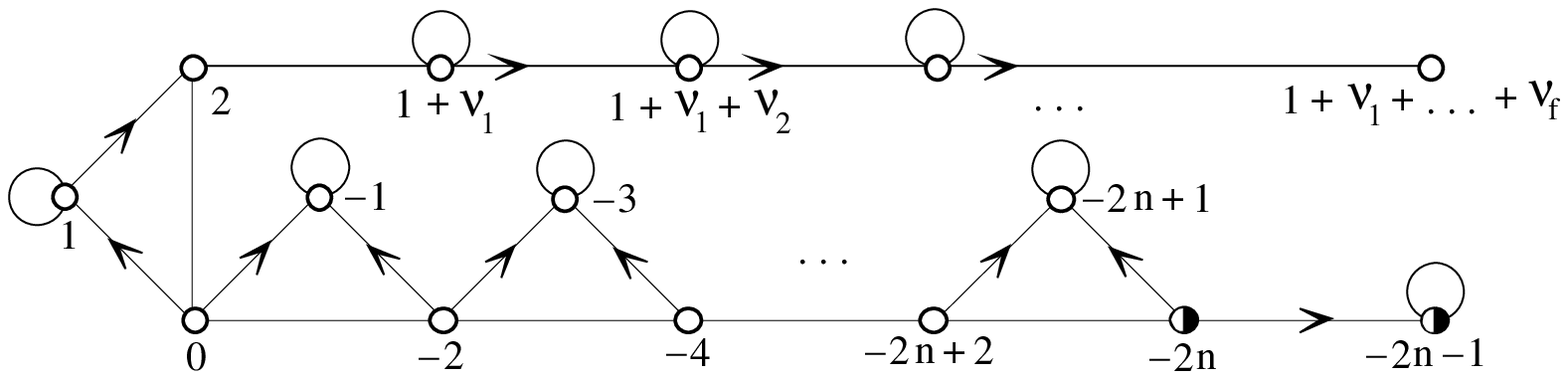}}
\nobreak
Fig.15 Incidence matrix ${\bf I}^{(1,5)}(p,p',n)$~\newinof~ for
the $\phi_{1,5}$ perturbation family of $M({\bar p},{\bar p}')$ 
for $2<{\bar p}'/{\bar p}<5/2.$ The relation between ${\bar p}, {\bar
p}'$ and $p,p',n$ is given in ~\ynew~and ~\another.

\listrefs

\vfill\eject

\bye
\end